\documentclass{emulateapj}
\usepackage{color,amsmath}
\newcommand{\fAGN}{\ensuremath{f({\rm AGN})_{\rm MIR}}}
\newcommand{\fTOT}{\ensuremath{f({\rm AGN})_{\rm IR}}}

\begin{document}

\title{CO Emission in Infrared-Selected Active Galactic Nuclei}
\author{Allison Kirkpatrick\altaffilmark{1,2}, Chelsea Sharon\altaffilmark{3,4}, Erica Keller\altaffilmark{5,6}, Alexandra Pope\altaffilmark{5}}
\altaffiltext{1}{Department of Physics \& Astronomy, University of Kansas, Lawrence, KS 66045, USA, akirkpatrick@ku.edu}
\altaffiltext{2}{Yale Center for Astronomy \& Astrophysics, New Haven, CT 06520, USA}
\altaffiltext{3}{Yale-NUS College, Singapore 138527} 
\altaffiltext{4}{Department of Physics \& Astronomy, McMaster University, Hamilton, Ontario L8S-4M1, Canada}
\altaffiltext{5}{Department of Astronomy, University of Massachusetts, Amherst, MA 01002, USA}
\altaffiltext{6}{National Radio Astronomy Observatory, Charlottesville, VA, 22903, USA}

\begin{abstract}
In order to better understand how active galactic nuclei (AGN) effect the interstellar media of their host galaxies, we perform a meta-analysis of the CO emission for a sample of $z=0.01-4$ galaxies from the literature with existing CO detections and well-constrained AGN contributions to the infrared (67 galaxies). Using either {\it Spitzer}/IRS mid-IR spectroscopy or {\it Spitzer}+{\it Herschel} colors 
we determine the fraction of the infrared luminosity in each galaxy that can be attributed to heating by the AGN or stars. We calculate new average CO spectral line ratios (primarily from \citealt{carilli2013}) to uniformly scale the higher-$J$ CO detections to the ground state and accurately determine our sample's molecular gas masses. We do not find significant differences in the gas depletion timescales/star formation efficiencies (SFEs) as a function of the mid-infrared AGN strength ($\fAGN$ or $L_{\rm IR} ({\rm AGN})$), which indicates that the presence of an IR-bright AGN is not a sufficient sign-post of galaxy quenching. We also find that the dust-to-gas ratio is consistent for all sources, regardless of AGN emission, redshift, or $L_{\rm IR}$, indicating that dust is likely a reliable tracer of gas mass for massive dusty galaxies (albeit with a large degree of scatter). Lastly, if we classify galaxies as either AGN or star formation dominated, we do not find a robust statistically significant difference between their CO excitation.

\end{abstract}

\section{Introduction}

The production of luminous quasars is a dramatic story, wherein two immense galaxies collide, fueling a burst of star formation and triggering rapid growth of a supermassive black hole \citep[e.g.,][]{sanders1996,hopkins2006,goulding2018}. In this scenario,
the active galactic nucleus (AGN) goes through an obscured growth phase, where the accretion disk is hidden by a dust torus. This phase ends when the AGN launches winds powerful enough to blow away some of the 
circumnuclear obscuring dust so that the accretion disk becomes visible in the optical \citep{glikman2012}.
As the winds clear away the circumnuclear dust, the galaxy's star formation quenches due to galaxy-scale, AGN-driven outflows \citep{page2012,pontzen2017}. However, evidence for this scenario is strongly linked to observational biases. In large statistical samples of optically-selected AGN \citep{stanley2017} and X-ray+IR selected AGN \citep{harrison2012}, no correlation between AGN luminosity and decreased star formation rate is observed. On the other hand, theoretical models suggest AGN may occur at a special phase in a galaxy's life, immediately prior to quenching \citep{caplar2018}. X-ray selected AGN at $z\sim1-2$ are observed to occur after a galaxy has compactified and but {\it before} star formation  decreases in the center of the galaxy \citep{kocevski2017}.

At $z>1$, galaxies both on and above the ``main sequence" \citep[e.\/g.\/,][]{brinchmann2004,daddi2007,elbaz2007,noeske2007} appear to have higher gas masses, gas mass fractions, and star formation rates than local galaxies \citep[e.\/g.\/,][]{tacconi2013,scoville2016}. The high star formation rates (SFRs) and gas masses of dusty galaxies are a particular challenge for simulation work, both in terms of reproducing the characteristics of these high-$z$ systems and in terms of the resulting ``red and dead" galaxy populations that would exist today (e.\/g.\/, \citealt{casey2014} and references therein; though see \citealt{narayanan2015}). Given the tight correlation observed between galaxies' black hole masses and bulge masses at $z\sim0$ (implied by the $M_{\rm BH}$-$\sigma$ relation; \citealt{ferrarese2000,gebhardt2000}) and the concurrent peaks of both AGN and star formation activity at $z\sim2$, theoretical studies have suggested that AGN may play a role in quenching the high SFRs of massive galaxies in the early universe \citep[e.\/g.\/,][]{somerville2008}.

However, AGN may not be solely responsible for quenching star formation. Recently, \citet{spilker2018} found molecular mass outflow rates a factor of two higher than the SFR in a star forming galaxy at $z=5.3$ that lacked an AGN. In contrast, \citet{hunt2018} find a quenched galaxy at $z\sim0.7$ with a large molecular gas reservoir and no indication of gas outflows. Indeed, post-starburst galaxies, which lack AGN signatures, are seen to retain large molecular gas reservoirs (albeit with depleted dense gas;  \citealt{french2015,suess2017,french2018}). How these results fit with growing supermassive black holes into a quenching paradigm remains unclear.
Both the relative importance of AGN feedback to stellar feedback \citep[e.\/g.\/,][]{bouche2010,dave2011a,shetty2012,lilly2013,kim2013} and exact feedback mechanism for AGN (outflows vs.~accretion suppression; e.\/g.\/, \citealt{croton2006,hopkins2006,gabor2011,cicone2014}) are still debated. In addition, recent work suggests that AGN may locally enhance star formation \citep[e.\/g.\/,][]{stacey2010,ishibashi2012,silk2013,mahoro2017}.

While the influence of AGN on galaxy-wide scales is unclear, AGN are known to heat their torus and circumnuclear dust to $>1000$\,K, producing an IR SED that emits as a power-law in the near and mid-IR \citep{elvis1994}. Empirically, dust-enshrouded AGN are observed to heat circumnuclear dust out to at least 40\,$\mu$m, producing a flattening in the far-IR \citep{mullaney2011,sajina2012,kirkpatrick2012,symeonidis2016,ricci2017}. Physical radiative transfer models of the narrow line region 
(NLR) also predict a modified blackbody peaking at $\lambda\sim30-100\mu$m from dust clouds comingled with the ionized gas \citep{groves2006}.
AGN may therefore have a non-negligible contribution to the infrared luminosity ($L_{\rm IR}$) and thus any inferred IR SFRs if the AGN is not properly accounted for. The far-IR influence of AGN produces an enhanced dust emission with temperatures of $\sim$100\,K \citep{kirkpatrick2012}. Gas and dust are largely considered to be comingled within the interstellar medium, at least locally, implying that AGN could be heating the molecular reservoirs in the centers of galaxies as well, producing a possible feedback mechanism that does not require outflows. However, at $z>1$, gas fractions have dramatically increased, and there has been some suggestion that the gas reservoirs are more spatially extended than the dust (e.\/g.\/, \citealt{hodge2015,hodge2016,tadaki2017b} but see also \citealt{hodge2018}).

If AGN influence the star formation history of galaxies, their effects may be measured in the molecular ISM that fuels star formation. Molecular outflows have been observed in the CO rotational lines of luminous Type 1 and Type 2 QSOs at low redshift \citep[e.\/g.\/,][]{feruglio2010,cicone2014,fiore2017}, but outflows are not present in all AGN host galaxies. Very high excitation CO lines ($J_{upper}\gtrsim10$) have been observed in some AGN at all redshifts \citep[e.\/g.\/,][]{weiss2007a,ao2008,hailey2012,spinoglio2012b,rosenberg2015}, but this component likely represents a small fraction of their total molecular gas reservoirs. At high redshift, the CO spectral line energy distributions (SLEDs) of Type 1 QSOs are known to peak, on average, at higher-$J$ transitions ($J_{upper}\sim7$) than those of star-forming submillimeter-selected galaxies (SMGs; $J_{upper}\sim5$; e.\/g.\/, \citealt{weiss2007b,carilli2013}). However, when high star formation rate densities can explain SLED excitation \citep[e.\/g.\/,][]{narayanan2014} it is challenging to isolate the potential effects of the AGN in driving these more moderate excitation CO rotational lines.

In this paper, we reanalyze data from the literature to examine whether AGN that are heating circumnuclear dust may also affect the molecular ISM. In our reanalysis, we demonstrate how different measurements of infrared luminosity and AGN heating can cause incongruent conclusions about the state of the ISM in AGN. In Section \ref{data}, we discuss our sample selection. In Section~\ref{methods}, we describe how we assess the presence and strength of obscured AGN, determine $L_{\rm IR}$ and the inferred SFRs, and use new average SLEDs from the literature to calibrate the gas masses of our sample. In Section \ref{discussion}, we look for evidence of decreased star formation efficiency in AGN and evaluate CO SLEDs for effects from the central AGNs. Finally, in Section \ref{conclusions}, we summarize and lay out future directions for uncovering the true effect of an AGN on its host ISM.

\section{Sample Selection}
\begin{figure}
    \centering
    \includegraphics[width=3.4in]{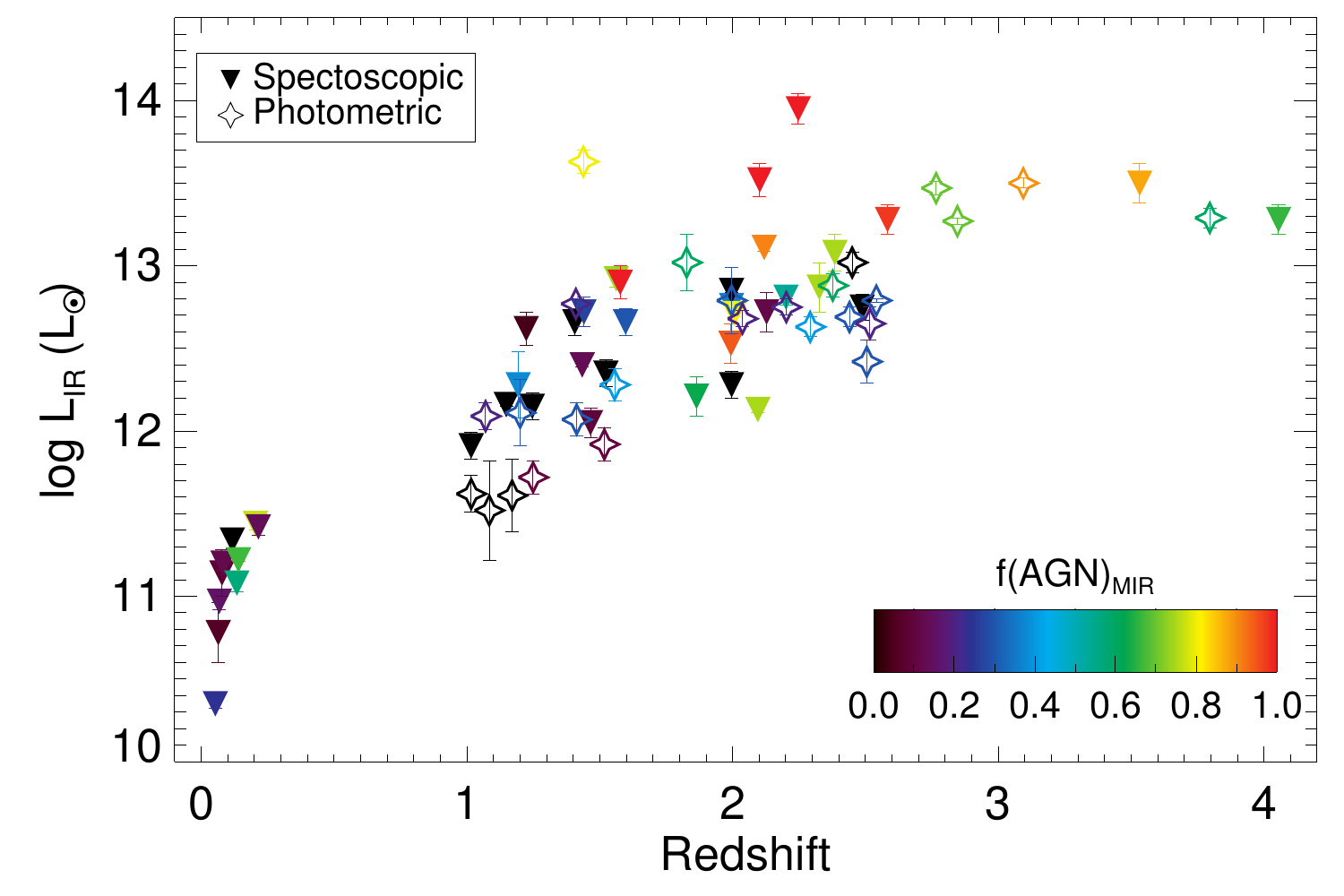}
    \caption{$L_{\rm IR}-z$ distribution of our sample. The $z<0.5$ sources are from \citet{kirkpatrick2014}, while the $z>1$ galaxies are predominantly mined from the \citet{carilli2013} compilation. We color sources according to their mid-IR AGN contribution.} The triangles indicate galaxies where we identified IR AGN using {\it Spitzer} mid-IR spectroscopy, while the stars are galaxies where AGN were found using IR color diagnostics.
    \label{lirz}
\end{figure}

We aim to compare the CO emission properties of IR-selected AGN with star forming galaxies. Due to the lack of a homogeneous survey, we reanalyze published sources. This technique has also been employed by other authors due to the difficulty of carrying out large scale molecular gas surveys \citep[e.g.][]{herrera2019,perna2018}. The meta-analysis necessarily means that our sample will be heterogeneous, although we apply the same analysis techniques uniformly to each source.
\label{data}
The relatively recent \citet{carilli2013} annual review described molecular gas in galaxies at $z>1$. This review compiled 173 galaxies with spectroscopic redshifts where molecular gas had been measured, comprising all published measurements at that time. From this parent sample, we initially selected all galaxies with a CO line flux for any $J$ transition, that have a magnification $\mu\leq5$, and with at least 3 photometric data points spanning the range $5-1000\,\mu$m. We also removed two galaxies (SMM\,J105131 and SMM\,J09431+4700) since they are resolved doubles in their CO emission, but we do not have matching resolved mid-IR colors or IRS spectroscopy to disentangle the AGN emission from the two sources.\footnote{SMM\,J123707+6214 is also a resolved double. However, the combined system has a measured mid-IR AGN fraction of $0\%$, so we do not have to worry about differences in the two components' AGN fractions. In addition, the CO line ratios are identical between the two components \citep{riechers2011c}, so we can treat the system as one for our analysis without any biasing concerns (this effectively under-weights SMM\,J123707+6214).} 
We also add in three 70\,$\mu$m selected galaxies (GN70.38, GN70.8, GN70.104) from \citet{pope2013}, where CO was observed with IRAM/PdBI; these galaxies were not in the \citet{carilli2013} compilation. We add in three more QSOs (COB011, no.226, 3C 298) from the \citet{perna2018} compilation which meet our criteria for full coverage of the IR SED. Additionally, we include 10 galaxies at $z<0.5$ from \citet{kirkpatrick2014}, which have CO(1-0) measured with the Large Millimeter Telescope, have {\it Herschel} observations covering the far-IR, and have {\it Spitzer} mid-IR spectroscopy. We have a total heterogenous sample of 67 galaxies. 
Figure \ref{lirz} shows the redshift-$L_{\rm IR}$ distribution of our final sample of 67 galaxies. We list all sources, redshifts, and \fAGN\ in Table \ref{properties} in the Appendix.

\section{Methods: AGN fraction, SFR, $M_{\rm dust}$, and $L_{\rm CO}^\prime$}
\label{methods}

\subsection{AGN fractions from Spitzer Spectroscopy and IR Colors}

We use either {\it Spitzer} IRS spectroscopy (available for 41 galaxies) or IR color diagnostics (26 galaxies) to locate AGN within our sample. We quantify \fAGN, the fraction of mid-IR ($5-15 \mu$m) luminosity arising from an AGN torus. We do not perform a full SED decomposition into multiple components as has become common in the literature because this risks overfitting and overinterpretting the sparse amount of IR data available.

The low resolution ($R=\lambda/\Delta \lambda \sim 100$) {\it Spitzer} IRS spectra were reduced following the method outlined in \citet{pope2008}. We used the {\it Spitzer} IRS spectra to calculate the AGN contribution to the mid-IR luminosity following the technique described in \citet{pope2008} and \citet{kirkpatrick2015}.
We fit the individual spectra with a
model comprised of three components: (1) the star formation component is represented by the mid-IR spectrum of the prototypical starburst M\,82;
(2) the AGN component is determined by fitting a pure power-law with the slope and normalization
as free parameters; (3) an extinction curve from the \citet{draine2003} dust models is applied to the AGN component; for our wavelength range, the extinction curve primarily constrains the 9.7\,$\mu$m silicate absorption feature. We fit all three components simultaneously.
We integrate under the best fit model and the power-law component to measure \fAGN. Our spectroscopic measurement technique is uncertain by less than 10\% in \fAGN\ \citep{kirkpatrick2015}.

 For the 26 sources that lack mid-IR spectroscopy, we determine \fAGN\ using {\it Spitzer}+{\it Herschel} color classification as described in \citet{kirkpatrick2017}, which is based on observed AGN emission in galaxies at $z\sim1-3$. We created a redshift dependent diagnostic that assigns an \fAGN\ depending on the ratio of $S_8/S_{3.6}$, which at $z>1$ separates AGN on the basis of a hot torus as compared with the 1.6\,$\mu$m stellar bump seen in star-forming galaxies (SFGs). This is combined with either $S_{250}/S_{24}$ or $S_{100}/S_{24}$, which trace the relative amounts of cold and hot dust (since AGN should have a larger hot dust component). Color determination of \fAGN\ in this manner is accurate to 30\% \citep{kirkpatrick2017}.

Our mid-IR classifications largely agree with the classifications in \citet{carilli2013}. That is, sources with $\fAGN<0.5$ are predominately listed as a type of star forming galaxy in \citet{carilli2013}, and the radio AGN and QSOs identified in \citet{carilli2013} all have $\fAGN\geq0.6$.
In addition, we identify 9 additional galaxies as AGN that were originally listed as star forming galaxies in \citet{carilli2013}. 
We note our 9 ``new'' AGN in Table \ref{properties}.

\subsection{$L_{\rm IR}$ and SFRs}

\begin{figure}
\includegraphics[width=3.3in]{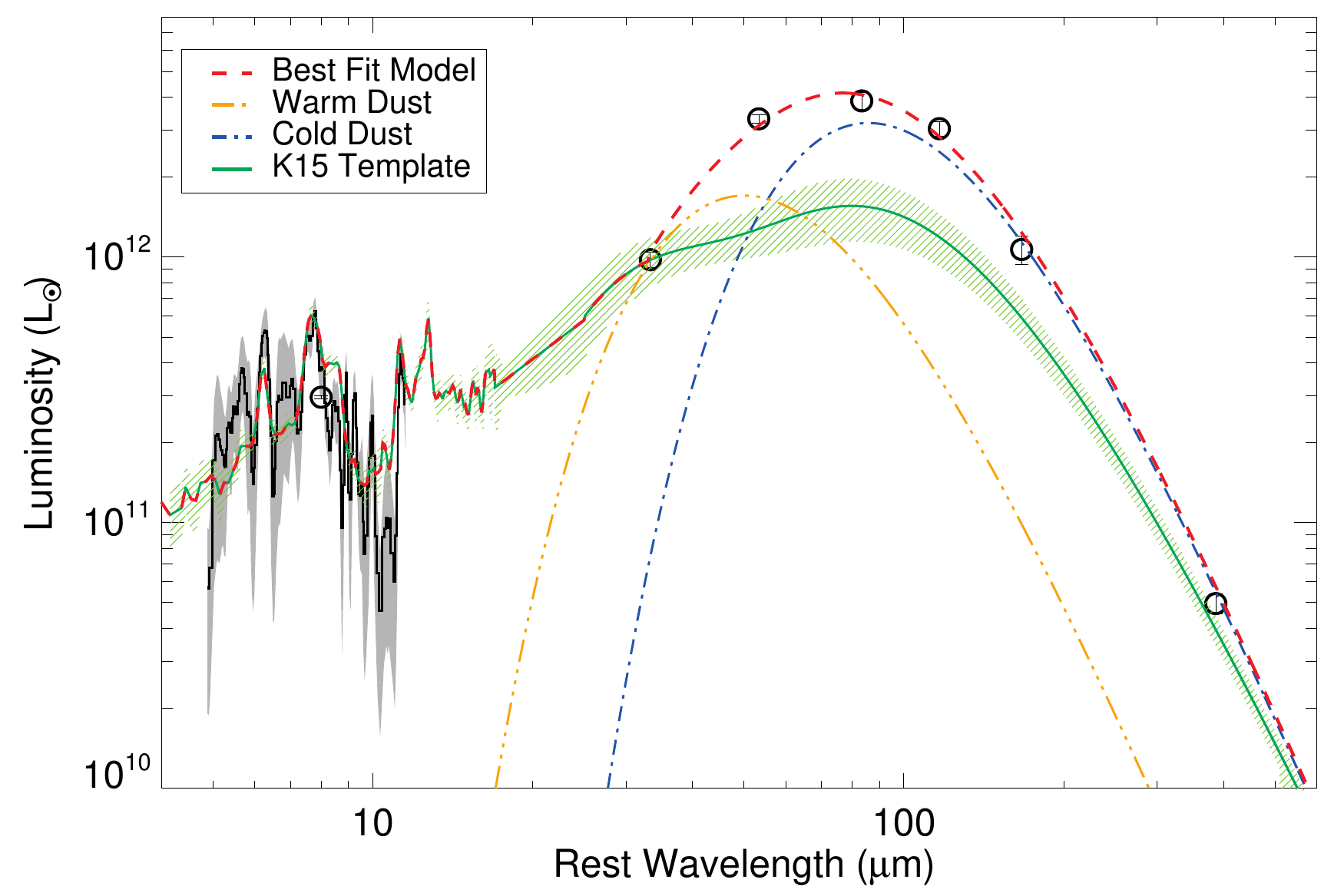}
\caption{Example showing our technique of calculating $L_{\rm IR}$ by combining a template (green solid line) from \citet{kirkpatrick2015} with a two-temperature modified blackbody model in the far-IR. The warm component is represented by the orange dot-dashed line and cold component by the blue dot-dashed line. The combined best-fit model shown with the red dashed line. This method more accurately determines $L_{\rm IR}$ than the template on it's own.\label{lir}}
\end{figure}

We gathered all available photometry for our sample using the {\it Spitzer}/IRSA and NED databases. The number of far-IR data points per source varies. To allow the greatest flexibility 
when determining $L_{\rm IR}$, we combine a template from \citet{kirkpatrick2015}, which are empirically derived from galaxies at $z\sim1-3$, with a far-IR model consisting of two modified blackbodies. The number of data points at $\lambda>20\,\mu$m determines how many far-IR parameters we 
allow to vary, so that we are at most fitting for $n-1$ 
parameters. At a minimum, we only fit for the normalization of the cold component, and a maximum, we fit both normalizations 
and the cold dust temperature. The warm dust temperature is held fixed at 65-90\,K, where using a warmer dust temperature depends on the presence of an AGN \citep{kirkpatrick2012}. Here, we use a warmer dust temperature when $\fAGN>0.6$, which is when this component starts to contribute measurably ($>20\%$) to the far-IR SED \citep{kirkpatrick2015}. 
We choose this simple method rather than 
using a library of models because we do not want to risk 
over-fitting a scarcity of data points in many sources. As we are not trying to measure dust temperatures or $\beta$, we are 
not concerned with degeneracies between the parameters. $L_{\rm IR}$ is robust against small variations in $T$ and $\beta$ as 
long as the model fits all the data points. The \citet{kirkpatrick2015} library provides templates based on 
\fAGN, and we use the appropriate template for each source to complete the best fit model at $\lambda \lesssim 20\mu$m. In Figure \ref{lir} we illustrate how our technique of modifying templates is more accurate than simply scaling a template to a photometric data point.

The $L_{\rm IR}$ can be converted to a SFR once the AGN heating is removed.
Due to the disparate data available for each source, we are not able to decompose the full SEDs into an AGN and star forming component. However
\citet{kirkpatrick2015} demonstrate that \fAGN\ is related to \fTOT, the AGN contribution to $L_{\rm IR}(8-1000\mu$m), through the quadratic relation
\begin{equation}
\label{eq:AGN}
\fTOT = 0.66 \times \fAGN^2 - 0.035\times \fAGN .
\end{equation}

\noindent
We correct $L_{\rm IR}$ for AGN heating using \fTOT using
\begin{equation}
L_{\rm IR}^{\rm SF} = L_{\rm IR} \times (1-\fTOT)
\end{equation}

\noindent
and then convert $L_{\rm IR}^{\rm SF}$ to a SFR following \citet{murphy2011}:
\begin{equation}
    {\rm SFR}\,[M_\odot {\rm yr}^{-1}] = (1.59\times10^{-10})\times L_{\rm IR}^{\rm SF}\,[L_\odot] .
\end{equation}

\subsection{Dust Masses}
\label{dust}

In order to explore whether increased AGN heating of the ISM affects the gas-to-dust ratio in galaxies, we use two methods for determining the dust mass in our sample.


For the first method, we calculate the dust mass and characteristic radius of the dust emission using the physically motivated, self-consistent radiative transfer model from \citet{chakrabarti2005,chakrabarti2008}. The far-IR SED can be characterized by the ratio of $L_{\rm IR}$ to $M_{\rm dust}$ and the radius of the source for a given dust opacity \citep{witt2000,misselt2001,chakrabarti2005,chakrabarti2008}. The authors use the \citet{weingartner2001} dust models with $R_V=5.5$ to parameterize the grain composition and emission properties. 
It should be noted that the choice of $R_V$ has a negligible effect in the far-IR, where $M_{\rm dust}$ is calculated. The authors make the simplifying assumption that star forming regions can be represented as homogeneous and spherically symmetric. Each star 
forming region is surrounded by spherical shells of dust, and the temperature and density profiles of the dust shells are approximated as power-laws. The radiative transfer code also calculates a far-IR luminosity ($\lambda > 30\mu$m), which is consistent with our $L_{\rm IR}$ calculated above. Many of our galaxies lack enough data to fit the near, mid, and far-IR with a robust dust 
model \citep[such as][]{draine2007} that calculates grain populations. As the relative abundance of small grains and carbonaceous 
grains does not have a significant effect on the large-grain submm emission, we opt for a model that only fits the far-IR. We list 
$M_{\rm dust}^{\rm CM}$ and $R$ from the \citet{chakrabarti2005} model in Table \ref{dat_table}.

\begin{figure}
    \centering
    \includegraphics[width=3.3in]{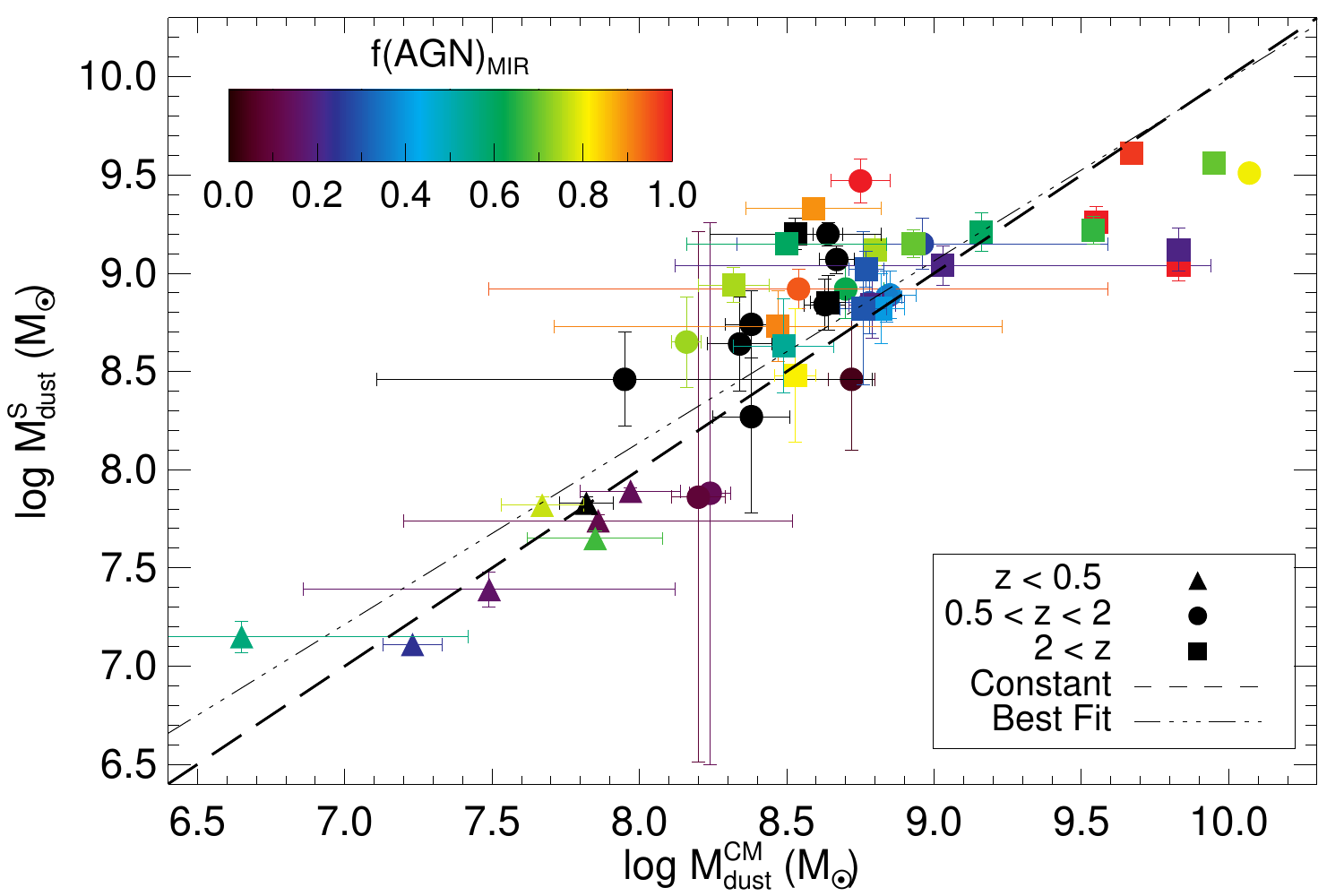}
    \caption{Comparison of dust mass derived from fitting \citet{chakrabarti2005} models on the x-axis with dust mass derived using the method in \citet{scoville2016} on the y-axis. The dashed line is the one-to-one correlation, while the dot-dashed line shows the best fit to the data using an ordinary least squares bisector method.}
    \label{fig:dust}
\end{figure}

For comparison to the \citet{chakrabarti2005} model, we also calculate the dust mass using the simple model of \citet{scoville2016}.
In this model, the 
dust mass is calculated using any sub-mm observation with $\lambda_{\rm rest}>250\,\mu$m, and assuming a standard cold dust 
temperature of 25\,K. Several of our galaxies lack a sub-mm observation at $\lambda_{\rm rest}>250\,\mu$m, so we are unable to calculate a dust mass in this manner.
We compare these dust masses, $M_{\rm dust}^{\rm S}$ and $M_{\rm dust}^{\rm CM}$, in Figure \ref{fig:dust}; their values are broadly consistent. The large errors on many $M_{\rm dust}^{\rm CM}$ indicate the unreliability of fitting a multi-component model to a scarce number of data points. Some of our galaxies only had two photometric measurements beyond $\lambda=30\,\mu$m.  

\subsection{$L^\prime_{\rm CO}$ and Gas Masses}
\label{gas}

\subsubsection{Excitation Correction}
\label{excitation}

Our \fAGN\ sample have been observed in many different CO lines. In order to determine $L^{\prime}_{\rm CO}$ and the associated gas masses free from excitation effects, assumptions need to be made about the line ratios in objects that do not have measurements of the \mbox{CO(1--0)} line (the ground-state rotational transition that best traces the total molecular gas). \citet{bothwell2013} produced an average CO spectral line energy distribution (SLED) for a sample of $z\sim1$--$4$ SMGs with spectroscopic redshifts (generally from Lyman-$\alpha$, ${\rm H\alpha}$, or PAH observations). However, there are significantly more CO detections currently in the literature, particularly more objects with \mbox{CO(1--0)} detections and more objects at higher redshifts.

We therefore construct a new average CO SLED using the {\it entire} library of high-$z$ CO detections from \citet{carilli2013}, supplemented with new CO detections from \citet{pope2013}, \citet{aravena2014}, \citet{sharon2016}, \citet{yang2017}, \citet{frayer2018}, \citet{perna2018} and references therein. We have included the entire library, as opposed to just those sources with well-sampled IR SEDs, in order to make use of all the available molecular gas information and therefore reduce uncertainties on our new SLED.

To construct the average CO SLED, we follow \citet{bothwell2013}. We linearly scale all sources' line fluxes and uncertainties by the ratio of their FIR luminosity to the approximate median FIR luminosity of the full sample ($5\times10^{11}\,L_\sun$) in order to prevent the brightest sources from biasing later averages. 
For each CO transition, we find the median scaled line luminosity for all sources detected in that line. We determine the uncertainty on the median scaled line luminosity via bootstrapping (with replacement) over 10,000 iterations, perturbing the random subset of line fluxes by their uncertainties in each iteration. We note that many sources from \citet{carilli2013} have small uncertainties. We assume that if those uncertainties are less than $10\%$ of the measured value, then only the statistical uncertainty was reported. For those objects, we add an additional $10\%$ uncertainty in quadrature to the reported value in order to approximate flux calibration uncertainties. For the very few line measurements with missing uncertainties, we assume a $20\%$ uncertainty. Note that this analysis only includes reported line \emph{detections}, which may produce a median SLED biased both towards the characteristics of the brighter classes of objects that are more commonly observed (which cannot be accounted for by luminosity scaling) and towards higher-excitation specifically (since upper limits from higher-$J$ non-detections are not included in the analysis). 

\begin{figure*}
\epsscale{1.1}
\plotone{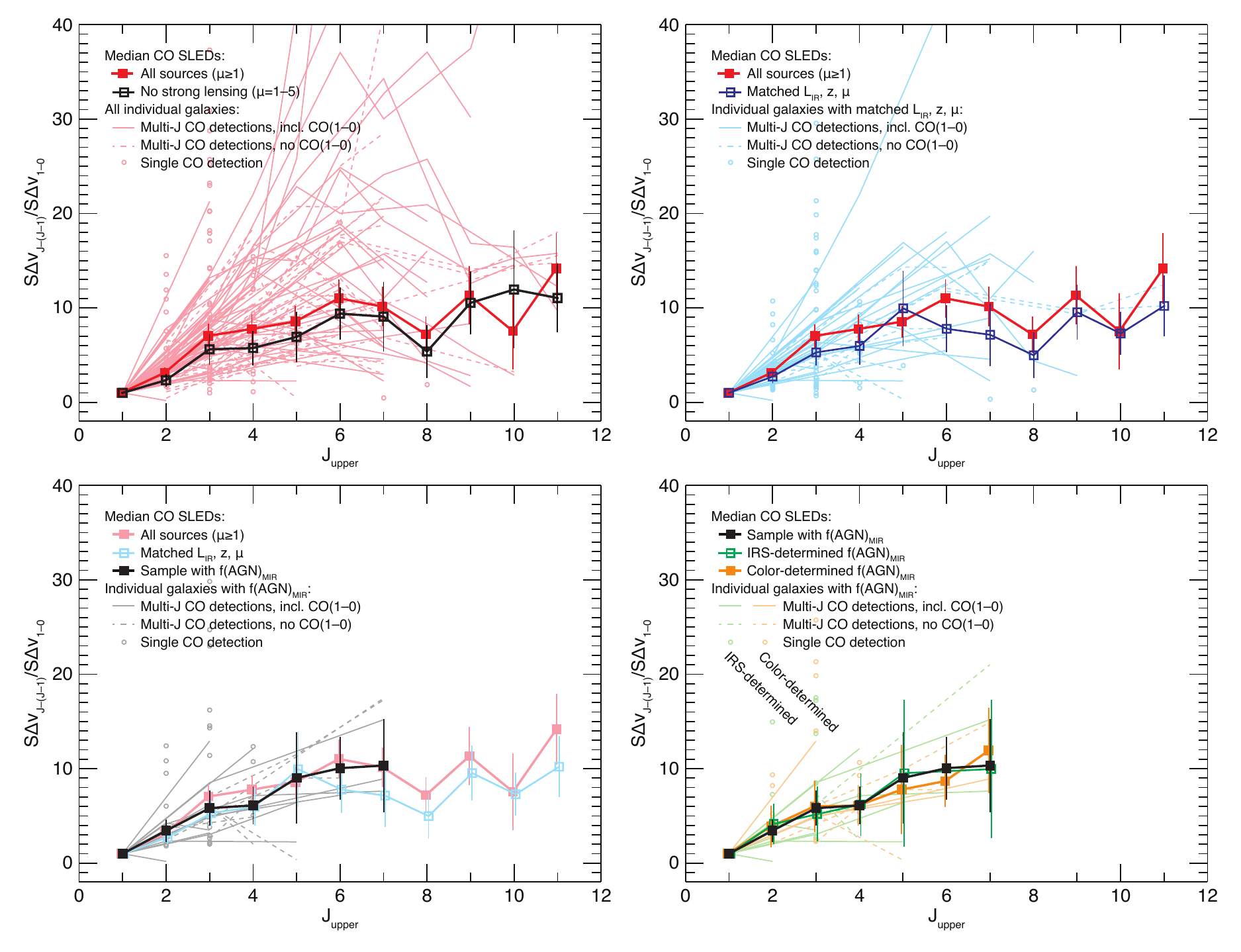}
\caption{Median CO SLEDs (squares and thick lines) and the SLEDs of the individual contributing galaxies (thin lines). Sources with only a single reported CO detection are shown as circles; the SLED of sources that lack published \mbox{CO(1--0)} measurements are shown as thin dashed lines. All sources that lack a \mbox{CO(1--0)} detections are scaled by the median \mbox{CO(1--0)} flux of the population as described in Section~\ref{gas}, and therefore their normalizations may change slightly from panel-to-panel (this does not affect the median CO SLEDs). {\it Top left:} the median CO SLED and line ratio uncertainties for the full sample from \citet{carilli2013} (red/solid squares; individual contributing sources in pink) and for the sub-sample with $1\leq\mu\leq5$ (black/open squares). {\it Top right:} we repeat the full sample's SLED (red/solid squares) for reference and show the median SLED for the sub-sample within the matching range of FIR luminosities, redshifts, and magnification factors as the IR galaxies of interest in this work (blue/open squares; individual contributing galaxies in light blue). {\it Bottom left:} we repeat the full sample's SLED (light red/solid squares) and matched sample's SLED (light blue/open squares), and show the median SLED for the galaxies with measured $f(AGN)_{MIR}$ (black/solid squares; individual contributing galaxies in gray). {\it Bottom right:} we repeat the median SLED for all galaxies with $f(AGN)_{MIR}$ (black/solid squares), and show the median SLEDs for the galaxies with IR AGN fractions determined from IRS spectra (green/open squares; individual contributing galaxies in light green) and the those determined from color templates (orange/solid squares; individual contributing galaxies in light orange). \label{fig:avesleds}}
\end{figure*}

We give the ratio of the median line luminosity to the median \mbox{CO(1--0)} line luminosity (using the scaled measured values) and the standard deviation of the ratio (from bootstrapping) in Table~\ref{tab:COratios}. The line ratios (in units of brightness temperature) can be related to the line luminosity and integrated line flux by

\begin{equation}
    r_{J,1}=\frac{L^\prime_{\rm CO(J\rightarrow J-1)}}{L^\prime_{\rm CO(1\rightarrow 0)}}=\frac{S\Delta v_{\rm CO(J\rightarrow J-1)}}{S\Delta v_{\rm CO(1\rightarrow 0)}}\left(\frac{1}{J}\right)^2
\end{equation}

\noindent
In Figure~\ref{fig:avesleds}, we plot the median CO SLED, as well as the individual line ratios contributing to that SLED. For individual sources that lack \mbox{CO(1--0)} measurements, we use the median \mbox{CO(1--0)} flux from our bootstrapped analysis, scaled by the ratio of that source's FIR luminosity to the fiducial value, in order to calculate and plot an illustrative line ratio for that object. 

\begin{deluxetable*}{@{\extracolsep{4pt}} ccccccccccccccccc}
\tablecaption{Median Brightness Temperature Ratios \label{tab:COratios}}
\tablehead{ \multicolumn{7}{c}{} & \multicolumn{6}{c}{Measured $f(AGN)_{\rm MIR}$ Sample} \\ \cline{8-13}
\colhead{} & \multicolumn{2}{c}{All Sources} & \multicolumn{2}{c}{$\mu\leq5$} & \multicolumn{2}{c}{Matched $L_{\rm IR}$, $z$, $\mu$} & \multicolumn{2}{c}{All} & \multicolumn{2}{c}{IRS-determined} & \multicolumn{2}{c}{Color-determined} \\ \cline{2-3} \cline{4-5} \cline{6-7} \cline{8-9} \cline{10-11} \cline{12-13}
\colhead{Transition} & \colhead{Ratio} & \colhead{$N$} & \colhead{Ratio} & \colhead{$N$} & \colhead{Ratio} & \colhead{$N$} & \colhead{Ratio\tablenotemark{a}} & \colhead{$N$} & \colhead{Ratio} & \colhead{$N$} & \colhead{Ratio} & \colhead{$N$}}
\startdata
CO(1--0) & \nodata & $69$ & \nodata & $41$ & \nodata & $37$ & \nodata & $16$ & \nodata & $11$ & \nodata & $5$\\
$r_{2,1}$ & $0.78\pm0.13$ & $73$ & $0.59\pm0.17$ & $55$ & $0.68\pm0.18$ & $38$ & $0.86\pm0.30$ & $21$ & $1.04\pm0.53$ & $15$ & $0.97\pm0.55$ & $6$ \\ 
$r_{3,1}$ & $0.78\pm0.14$ & $108$ & $0.62\pm0.18$ & $78$ & $0.59\pm0.15$ & $75$ & $0.65\pm0.21$ & $33$ & $0.58\pm0.32$ & $19$ & $0.67\pm0.31$ & $14$ \\  
$r_{4,1}$ & $0.49\pm0.10$ &  $54$ & $0.36\pm0.11$ & $41$ & $0.37\pm0.12$ & $29$ & $0.38\pm0.12$ & $12$ & $0.39\pm0.21$ & $5$ & $0.38\pm0.13$ & $7$ \\ 
$r_{5,1}$ & $0.34\pm0.07$ &  $41$ & $0.28\pm0.11$ & $24$ & $0.40\pm0.16$ & $11$ & $0.36\pm0.19$ & $5$ & $0.38\pm0.31$ & $2$ & $0.31\pm0.19$ & $3$ \\ 
$r_{6,1}$ & $0.31\pm0.06$ &  $47$ & $0.26\pm0.08$ & $32$ & $0.22\pm0.07$ & $11$ & $0.28\pm0.09$ & $2$ & \nodata & $0$ & $0.24\pm0.08$ & $2$ \\ 
$r_{7,1}$ & $0.21\pm0.04$ &  $37$ & $0.19\pm0.08$ & $21$ & $0.15\pm0.07$ & $11$ & $0.21\pm0.10$ & $5$ & $0.20\pm0.15$ & $3$ & $0.24\pm0.09$ & $2$ \\ 
$r_{8,1}$ & $0.11\pm0.03$ &  $13$ & $0.08\pm0.04$ & $5$ & $0.08\pm0.04$ & $5$ & \nodata & $0$ & \nodata & $0$ & \nodata & $0$\\ 
$r_{9,1}$ & $0.14\pm0.04$ &  $12$ & $0.13\pm0.04$ & $5$ & $0.12\pm0.04$ & $4$ & \nodata & $0$ & \nodata & $0$ & \nodata & $0$\\ 
$r_{10,1}$ & $0.08\pm0.04$ &  $7$ & $0.12\pm0.06$ & $2$ & $0.07\pm0.02$ & $1$ & \nodata & $0$ & \nodata & $0$ & \nodata & $0$\\ 
$r_{11,1}$ & $0.12\pm0.03$ &  $7$ & $0.09\pm0.03$ & $3$ & $0.08\pm0.03$ & $3$ & \nodata & $0$ & \nodata & $0$ & \nodata & $0$\\ 
\enddata
\tablecomments{The first three sets of median line ratios are calculated from the parent sample of \citet{carilli2013}, \citet{pope2013}, \citet{aravena2014}, \citet{sharon2016}, \citet{yang2017}, \citet{frayer2018}, \citet{perna2018} and references therein (which includes our sample with calculated mid-IR AGN fractions), and various subsets as indicated by the columns' titles. Median line ratios for our sample with measured mid-IR AGN fractions are shown in the last three columns. In all cases, $N$ refers to the number of galaxies with detections of the specified CO transition that contribute to the median line luminosity, and thus the median line ratio.}
\tablenotetext{a}{The astute reader may notice that combined \fAGN\ median ratios do not always fall between the color-determined and IRS-determined median ratios (similar effects are also seen for other ratios presented in this paper). This result is due to our method of calculating {\it a ratio of medians} as opposed to {\it a median of ratios} as described in Section~\ref{excitation} (which is necessary since not all sources have detections of both the relevant lines and we want to use all available information). Since we calculate the median $L^\prime_{\rm CO(J-(J-1))}$ and the median $L^\prime_{\rm CO(1-0)}$, those luminosities may skew in different directions depending on the number of detections and luminosities of their sub-populations. Thus the resulting ratio of median luminosities may be higher or lower than that of the objects or sub-samples from which the ratio is composed.}
\end{deluxetable*}

Many high-$z$ sources studied in CO are gravitationally lensed due to the advantageous boosting of the observed flux. However, spatially varying magnification factors may interact with spatially varying properties of the galaxy, biasing unresolved measurements of lensed galaxies' fluxes \citep[e.\/g.\/,][]{blain1999,hezaveh2012,serjeant2012}. In a series of simulations, \citet{serjeant2012} showed that differential lensing tends to scatter unresolved measurements of galaxies' average properties towards those of their most compact and luminous regions (like those of the central AGN and star-forming clouds, which are typically warmer than the ambient material), particularly for strongly lensed sources with magnifications of $\mu>10$. Since differential lensing could bias measured CO SLEDs towards more high-$J$ CO emission, we also compute the median CO SLED as above, but restrict the included sources to those with magnifications of $\mu\leq5$, where \citet{serjeant2012} found little evidence for differential lensing affecting the shape of galaxy-wide average CO SLEDs (Figure~\ref{fig:avesleds}, top left). This selection reduces the number of lines in the sample by $\sim1/3$. While removing the strongly-lensed $\mu>5$ sources does systematically produce lower line ratios (at least where there are a significant number of input line measurements), differences in individual line ratios are not statistically significant (Table~\ref{tab:COratios}).

Galaxies' observed CO SLEDs can also be affected by the cosmic microwave background (CMB), which can both act as an additional heating mechanism and as a background above which the CO lines must be detected \citep[e.\/g.\/,][and references therein]{dacunha2013}. For galaxies at higher redshifts ($z\gtrsim5$), \citet{dacunha2013} finds that the CMB's effect on theoretical CO SLEDs can be substantial. However, the effects on the measured line ratios also depend on the density and temperature of the molecular gas. We therefore cannot apply redshift-dependent corrections to individual galaxies in our sample since we do not have {\it a priori} constraints on their gas physical conditions. Therefore, our redshift-binned samples will have some additional scatter introduced by the effects of the CMB. However, the effects of the CMB in the redshift range of interest ($1\leq z\leq 4$) are generally much smaller than the scatter introduced by the galaxies' intrinsic range of gas temperatures and densities \citep{dacunha2013}. 

To ensure that including such a broad range of redshifts and IR luminosities will not bias our updated SLED, we also calculate the median line ratios for the sub-sample of galaxies from \citet{carilli2013}, \citet{perna2018}, etc. that matches the redshift ($1\leq z\leq 4$), magnification ($\mu\leq5$), and IR luminosity ($0.1\times10^{12}\,{\rm L_\sun}\leq L_{\rm IR}\leq90\times10^{12}\,{\rm L_\sun}$) of our mid-IR sample described in Section \ref{data}. The median CO line ratios are reported in Table~\ref{tab:COratios} and the CO SLED is plotted in Figure~\ref{fig:avesleds} (top right). 
There is no significant difference between the median line ratios for this 
sub-sample and the full literature sample. 

Finally, we compare how the CO excitation of our IR selected sample compares with the CO SLED using the full literature sample.
In Figure \ref{fig:avesleds}, we show the median CO SLED for all sources with \fAGN (bottom left panel), and consider the sub-samples with IRS-determined \fAGN\ and color-determined \fAGN\ separately (bottom right panel; listed in Table~\ref{tab:COratios}). The sample with measured MIR AGN fractions has only $\sim1$--30 CO detections per line, and no CO observations above the \mbox{CO(7--6)} transition. Again, we find no significant difference between the median CO line ratios for the \fAGN\ sub-sample and the complete literature sample. We consider the effect of AGN on the SLED in Section \ref{sec:agn_sled}.

Though there is no statistically significant difference between the line ratios for the various median CO SLEDs, we use the median ratios calculated from the low-magnification sub-sample that matches the redshift and IR luminosities of the sample of interest (top right of Figure \ref{fig:avesleds}) to avoid any potential biases when extrapolating down from higher-$J$ CO lines to the \mbox{CO(1--0)} line luminosity. The highest-$J$ transition we use to determine the ground state CO line luminosity is the \mbox{CO(4--3)} line. We use these excitation-corrected CO line luminosities to determine molecular gas masses.

\subsubsection{Choice of CO-to-${\rm H_2}$ Conversion Factor}
\label{sec:alpha}

Calculating gas mass from $L^\prime_{\rm CO}$ requires assuming a conversion factor, $\alpha_{\rm CO}$, that depends on the conditions in the ISM. These assumptions range from using a single conversion factor for all galaxies \citep[e.g.][]{scoville2016}, to a bimodal conversion for main sequence and starbursts \citep[e.g.][]{bolatto2013}, to a continuous conversion factor that depends on CO surface density and metallicity \citep[e.\/g.\/,][]{narayanan2012}. We explore two extremes: calculating gas mass with a single conversion factor and using a physically motivated continuous conversion factor from \citet{narayanan2012}.

The continuous conversion factor requires us to know the metallicity and the surface area of the ISM in our galaxies. For this literature-based sample, most CO observations are unresolved. However, the shape of the far-IR SED contains information about the physical extent of the dust emission; at the most fundamental level, the luminosity of a blackbody depends solely on temperature and radius. We make the simplifying assumption that the dust and gas are predominantly cospatial (though the few existing high resolution dust studies may challenge that assumption; e.\/g.\/, \citealt{hodge2015,hodge2016,tadaki2017b}; but see also \citealt{hodge2018}). We use the radius output by the \citet{chakrabarti2005} code to calculate surface area.

The continuous $\alpha_{\rm CO}$ conversion factor is given by

\begin{equation}
    \alpha_{\rm CO} = \frac{0.7*\Sigma_{\rm CO}^{-0.32}}{Z^{0.65}} 
\end{equation}

\noindent
where $\Sigma_{\rm CO}$ is the CO surface density in K km s$^{-1}$ and $Z$ is the metallicity \citep{narayanan2012}. To calculate the CO surface density, we use the characteristic radius from the far-IR fitting code (listed in Table \ref{dat_table}) and calculate an average surface density given by $\Sigma_{\rm CO}=L_{CO}^\prime/ \pi R^2$. 
Since we lack stellar masses for many galaxies, we cannot explicitly use the mass-metallicity relationship to estimate metallicities. However, $M_{\rm dust}$ also correlates with metallicity, so we follow the calibration in \citet{remy2015}:

\begin{equation}
\begin{aligned}
   \log(O/H)+12.0 &=\frac{\log M_{\rm dust}+15.0}{2.6}\\
Z &= 10^{(\log(O/H)+12.0)-8.69}
\end{aligned}
\end{equation}

\noindent
(where 8.69 is the value for solar metallicity). For the above expression, we use $M_{\rm dust}^{\rm CM}$ (rather than $M_{\rm dust}^{\rm S}$) to be consistent with our calculated radius.
This method yields CO-to-${\rm H_2}$ conversion factors of $0.08<\alpha_{\rm CO}<0.73$ (in units of $\rm{M_\sun (K\,km\,s^{-1}\,pc^2 )^{-1}}$; Table~\ref{dat_table}) and a mean value of $0.25\,\rm{M_\sun (K\,km\,s^{-1}\,pc^2 )^{-1}}$. All of the conversion factors calculated in this manner are less than the $\alpha_{\rm CO}=0.8$ determined for local dusty starbursts \citep{solomon1997} and frequently applied to high-$z$ starbursts. The relatively low conversion factors are likely a product of the compact dust sizes we use to calculate the CO surface brightness. However, without spatially resolved CO observations, the size of the dust emitting region the best estimate for the CO size currently available for this sample. See Section \ref{sec:gdr} for further discussion.

As a simple upper limit on the gas masses, we also assume a constant conversion factor of $\alpha_{\rm CO}=4.6$ \citep{solomon1991}, which is standard for main sequence galaxies. We do not know if our galaxies are on the main sequence as defined by SFR/$M_\ast$, since we do not have uniform $M_\ast$ measurements. Just based on $L_{\rm IR}$, \citet{sargent2012} find that at $z\sim1$, 90\% of galaxies with $L_{\rm IR}<3\times10^{11}\,L_\odot$ are on the main sequence, and 50\% of galaxies with $L_{\rm IR}<2\times10^{12}\,L_\odot$ are on the main sequence. Similarly, at $z=2$, 90\% of galaxies with $L_{\rm IR}<1.5\times10^{12}\,L_\odot$ are on the main sequence, and 50\% of galaxies with $L_{\rm IR}<8\times10^{12}\,L_\odot$ are on the main sequence. Given these luminosity estimates, our sample is likely a mix of main sequence and starbursting galaxies. Since we cannot distinguish between these for this sample, we use the main sequence conversion factor, although any observed trends would be unchanged if we instead used $\alpha_{\rm CO}=0.8$. We calculate $M_{\rm gas}^{\rm V}$ with the varying $\alpha_{\rm CO}$ and $M_{\rm gas}^{\rm MS}$ with $\alpha_{\rm CO}=4.6$ and list both in Table \ref{dat_table} in the Appendix.

\begin{figure*}
    \centering
    \includegraphics[width=\linewidth]{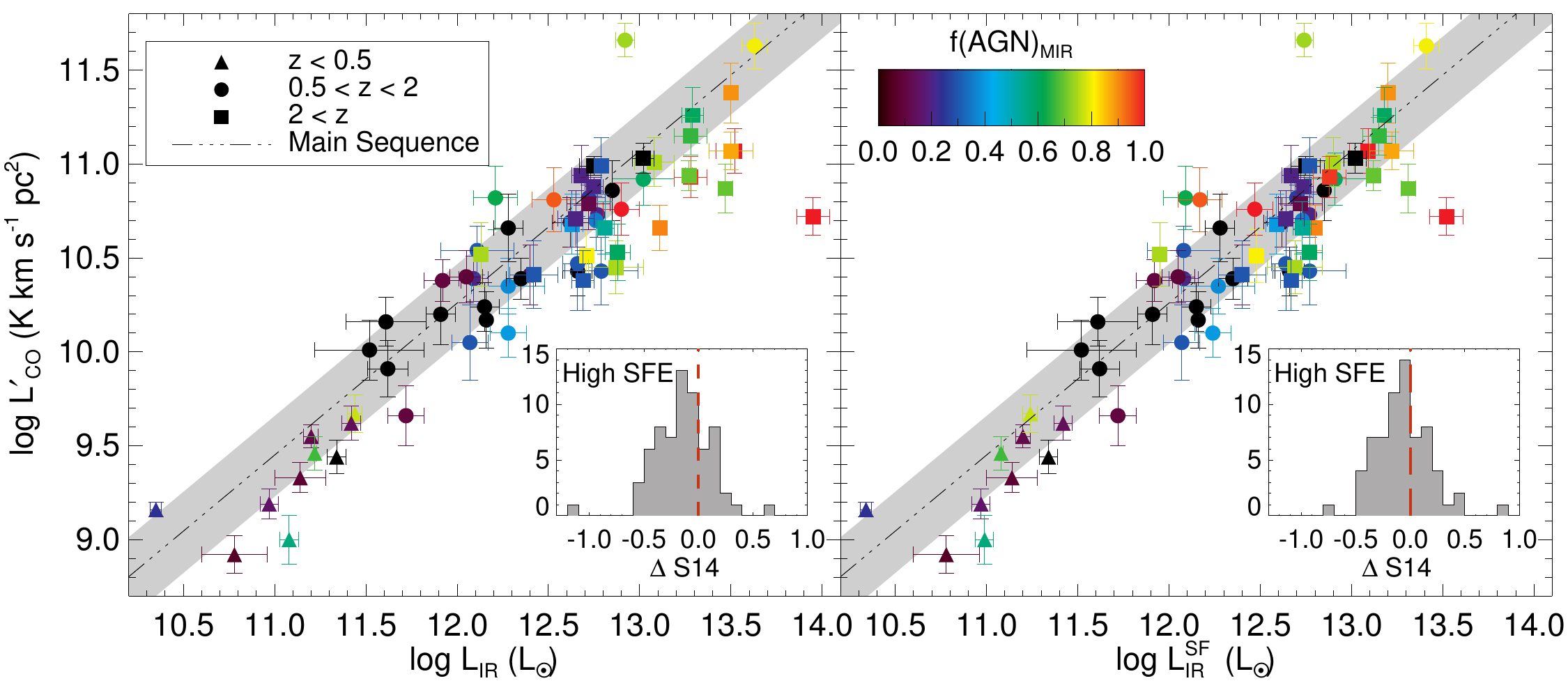}
    \caption{{\it Left:} Plot of $L^\prime_{\rm CO}$ vs.~$L_{\rm IR}$ for our sample without corrections for the AGN contributions to $L_{\rm IR}$. Symbols correspond to redshift as listed in the legend; points are colored by their \fAGN. The dashed line (shaded region) shows the relationship (1$\sigma$ scatter) derived in \citet{sargent2014} for normal star-forming galaxies at $z=0$--$3$. {\it Right:} Same as the left, except now we have removed the AGN contribution to $L_{\rm IR}$. This correction results in many of the $z>2$ galaxies lying closer to the relationship for ``normal" galaxies. In the insets, we show the distance of the galaxies from the \citet{sargent2014} relationship ($\Delta$S14), where high SFE occurs to the left of the dashed line. Before correcting for AGN heating, the mean distance from the normal relation is $-0.14\pm0.03$, whereas after the correction, the mean is $-0.07\pm0.03$---a small but statistically significant shift.}
    \label{lir_lco}
\end{figure*}

\section{Analysis \& Discussion}
\label{discussion}

\subsection{Star Formation Efficiency in AGN}
\label{sec:sfe}

In the past, galaxies have conventionally been categorized in terms of two different modes of star formation: a main sequence-like ``normal" mode (historically associated with local disk galaxies) and a starburst mode (historically associated with local U/LIRGs and frequently extended to high-$z$ SMGs). Normal star-forming galaxies have long gas depletion timescales of $>100\,$Myr (or for the depletion timescale's inverse: high star formations efficiencies; SFEs), while starbursts are compact and have short gas depletion timescales. However, due to increased gas fractions, the relationship between compactness, main sequence, starburst, and SFE becomes increasingly murky with lookback time \citep[e.\/g.\/,][]{kartaltepe2012,chang2018}. For example, in a sample of $z<1$ IR selected galaxies, \citet{kirkpatrick2014} demonstrate that the SFR/$M_\ast$ criteria does not correlate with star formation efficiency. In addition, typical assumptions of bimodal CO-to-${\rm H_2}$ conversion factors may induce a separation between normal galaxies and starbursts \citep[e.g.][]{daddi2010}, and hinges on distinguishing reliably between a starburst galaxy and a normal galaxy, rather than assuming a smooth transition between the two regimes. In order to avoid the complicating choice of conversion factor, \citet{carilli2013} use a standard convention of comparing $L_{\rm IR}$ with $L^\prime_{\rm CO}$ as a proxy for SFE, where starbursting galaxies have higher star formation efficiencies manifested as an elevated $L^\prime_{\rm CO} (\propto M_{\rm gas})$ for a given $L_{\rm IR}$ ($\propto$ SFR) \citep[see also][]{daddi2010,genzel2010,sargent2014}. In \citet{carilli2013}, radio AGN and QSOs had higher $L_{\rm IR}/L_{\rm CO}^\prime$ than main sequence selected galaxies.

In light of our recalculated homogeneous $L_{\rm IR}$ and $L^\prime_{\rm CO}$ values, and our quantitative/continuous approach to AGN emission, we revisit the relationship between $L_{\rm IR}$ and $L^\prime_{\rm CO}$ in Figure \ref{lir_lco}. We compare our sample to the $L_{\rm IR}$-$L^\prime_{\rm CO}$ correlation for normal star formation in $z=0-3$ galaxies from \citet{sargent2014}. We should take care here to note that this relationship is not a main sequence of star formation in the way that the relationship between SFR and $M_{\ast}$ is \citep[e.g.][]{noeske2007}. Rather, it is the region of parameter space where most main sequence galaxies (the so-called normal population) tend to lie. But a galaxy in the grey regime in Figure \ref{lir_lco} may not necessarily be a main sequence galaxy according to SFR/$M_\ast$.
Many of our $z>2$ galaxies lie off this norma relationship in the area of the graph (lower right) that would indicate a high SFE. This is not surprising 
since high-$z$ observations tend to be biased towards the brightest systems.
As the relationship between $L^\prime_{\rm CO}$ and $L_{\rm IR}$ is meant to be a proxy for the relationship between molecular gas mass and SFR (i.\/e.\/ an integrated form of the Schmidt-Kennicutt relation; \citealt{schmidt1959,kennicutt1989}), we should only be using the portion of $L_{\rm IR}$ due to star formation, $L_{\rm IR}^{\rm SF}$. We therefore show $L_{\rm IR}^{\rm SF}$ vs. $L^\prime_{\rm CO}$ in the right panel of Figure \ref{lir_lco}. Using $L_{\rm IR}^{\rm SF}$, almost every galaxy lies around the normal relation, which we quantify by calculating the difference in observed $L_{\rm CO}^\prime$ from the \citet{sargent2014} prediction for normal star formation given a galaxy's $L_{\rm IR}^{\rm SF}$. Before the AGN correction, the geometric mean difference ($\log L_{\rm CO}^\prime -\log L_{\rm CO}^\prime ({\rm S14})$) is $-0.14\pm0.03$, where the negative indicates that sources are underluminous in CO, corresponding to a higher SFE. After the AGN correction, the mean difference is $-0.07\pm0.03$; although it is small, this shift indicates how the presence of an AGN can bias interpretations of galaxy relationships if not properly accounted for.

\begin{figure}
    \centering
    \includegraphics[width=3.4in]{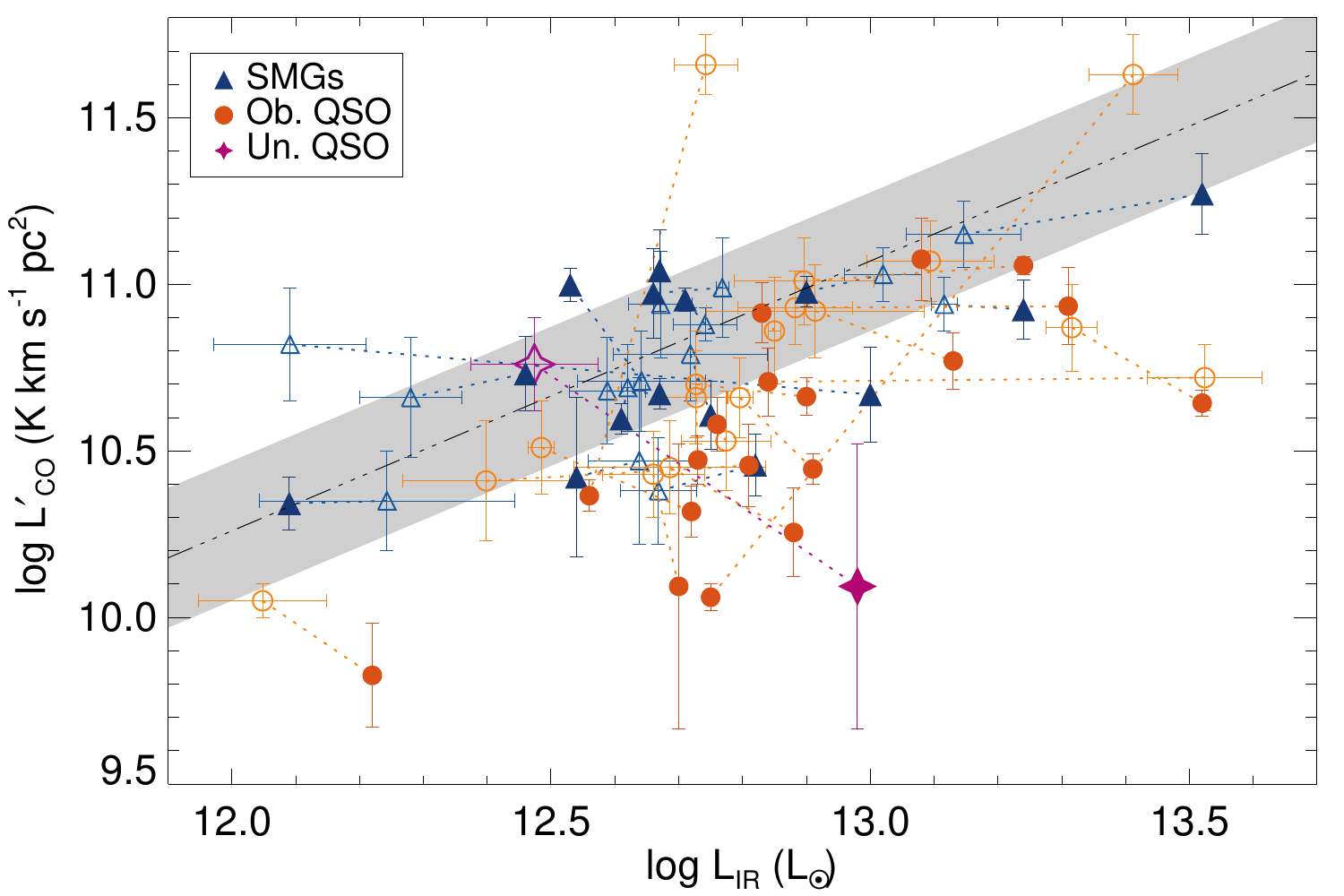}
    \caption{
    Comparison of the \citet{perna2018} measurements for SMGs (filled blue triangles), obscured quasars (filled orange circles) and unobscured quasars (filled purple star) with our measurements for those same sources (unfilled symbols, connected with dotted lines) on the $L_{\rm IR}$-$L_{\rm CO}^\prime$ relation. We have remeasured $L_{\rm IR}$ using high $z$ templates and accounting for AGN emission, and we have created new CO SLEDs to calculate $L_{\rm CO}^\prime$ from excited transitions. The different measurements can cause different interpretations of the role AGN play in galaxy evolution. The dashed line and shaded region is from \citet{sargent2014} as in Figure~\ref{lir_lco}. \label{perna}}
    \label{fig:my_label}
\end{figure}

 This result is noticeably differs from the recent work of \citet{perna2018} which finds that obscured and unobscured QSOs from \citet{carilli2013} have higher SFEs. Of their sample, 30 submillimeter galaxies, 35 obscured quasars, and 6 unobscured quasars meet our selection criteria of being unlensed, at $z<4$, and having a CO detection. Of these, only 15 SMGs, 18 obscured QSOs, and 1 unobscured QSO meet our criteria of having enough publicly available IR photometry to fully sample the SED. (All except 3 of these galaxies are drawn from the \citet{carilli2013} compilation.) \citet{perna2018} determine $L_{\rm IR}$ using a spectral decomposition method that relies on the \citet{chary2001} and \citet{dale2002} templates derived from local galaxies. As we demonstrated in \citet{kirkpatrick2012}, IR luminous galaxies at $z>1$ have significantly different far-IR emission than the \citet{chary2001} library and require their own set of templates. In Figure \ref{perna}, we show that the measurements from \citet{perna2018} do indeed place QSOs in the region of high star formation efficiency. However, when calculating $L_{\rm IR}$ using high-$z$ templates that account for AGN emission and calculating $L^\prime_{\rm CO}$ using our new CO SLEDs, the same QSOs lie much closer to the main sequence. This result clearly illustrates that conclusions regarding how galaxies evolve critically depend on how the measurements are made.

\begin{figure*}
    \centering
    \includegraphics[width=6.5in]{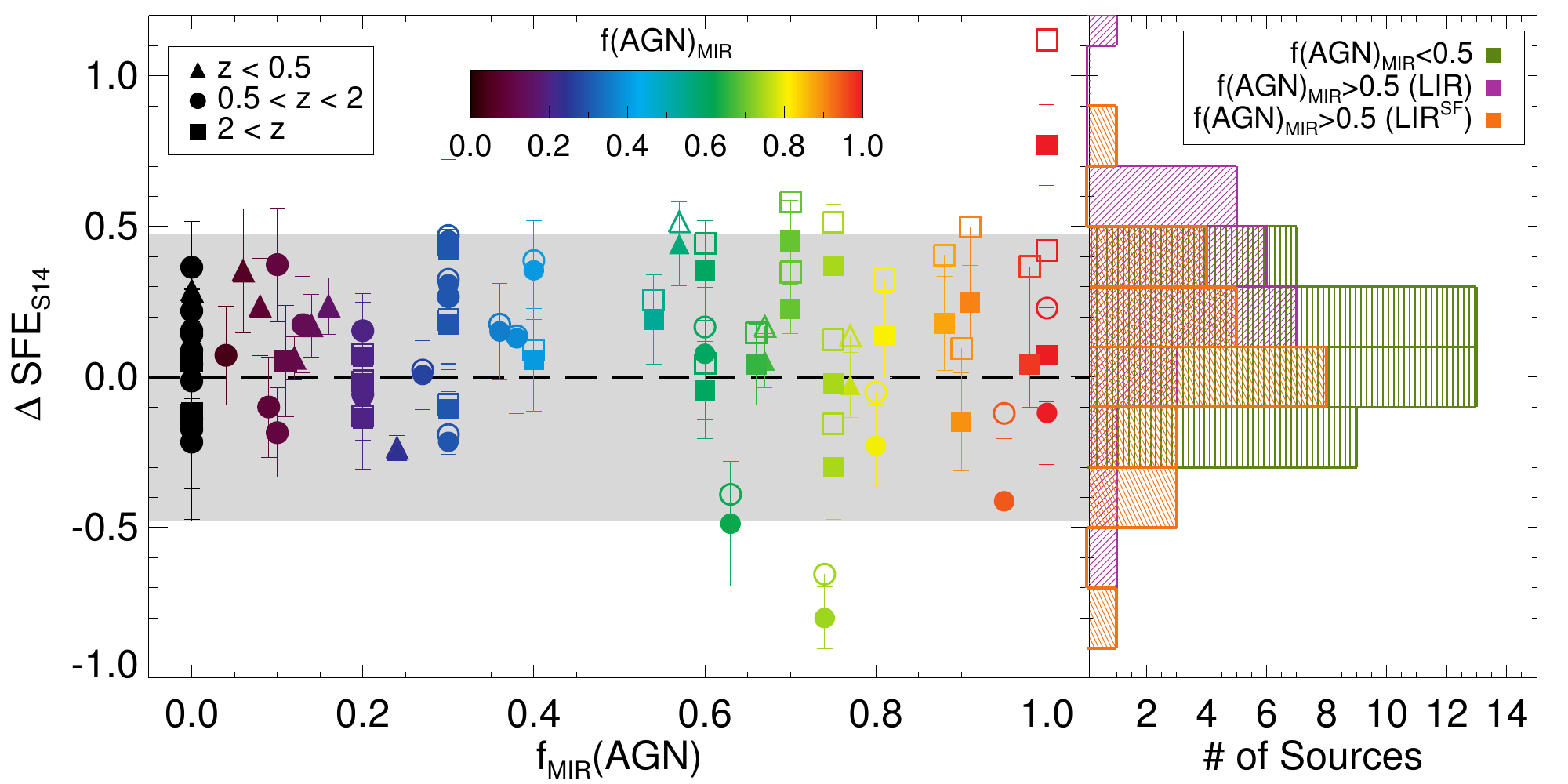}
    \caption{We look for a possible correlation between SFE (as parameterized by $L_{\rm IR}^{\rm SF}/L_{\rm CO}^\prime$) and AGN strength. We calculate the difference in SFE from the relationship measure by \cite{sargent2014}, shown in Figure \ref{lir_lco}, and plot $\Delta{\rm SFE}_{\rm S14}$ vs $\fAGN$ in the left panel. The shaded region represents a factor of 3 difference from the \citet{sargent2014} relationship. The right panel shows the distribution of galaxies with $\fAGN<0.5$ (green) and $\fAGN\geq0.5$ (orange). There is no trend of increasing or decreasing SFE with increasing \fAGN. If we had not corrected $L_{\rm IR}$ for AGN emission, there {\it would} be a trend, as exhibited by the unfilled symbols in the left panel and the purple histogram in the right panel.} 
    \label{SFE}
\end{figure*}

In Figure \ref{SFE}, we look for a correlation between black hole heating of the ISM and depressed star formation efficiency. We parameterize SFE as $L_{\rm IR}^{\rm SF}/L_{\rm CO}^\prime$. We first use the \citet{sargent2014} relation in Figure 5 to determine the SFE of normal galaxies. Then, we calculate the distance of our galaxies from this relation, $\Delta {\rm SFE}_{\rm S14}$. We plot $\Delta {\rm SFE}_{\rm S14}$ as a function of \fAGN\ in Figure \ref{SFE}. The shaded region represents a factor of 3 above and below the \citet{sargent2014} SFE. 
There is no trend between $\Delta {\rm SFE}_{\rm S14}$ and \fAGN\, as confirmed by a Spearman's rank test ($\rho=-0.02$, significance=0.85, which is not a statistically significant deviation from 0.0).

We split the sample at $\fAGN=0.5$ to measure whether low \fAGN\ galaxies have a different distribution of $\Delta {\rm SFE}_{\rm S14}$ than high \fAGN\ galaxies. We plot the distributions in the right panel of Figure \ref{SFE}. A two-population Kolmogorov-Smirnov test cannot rule out the null hypothesis that $\Delta {\rm SFE}_{\rm S14}$ of the two-subsamples are drawn from the same parent population. In other words, for our limited, heterogenous sample, there is no significant difference in the SFEs of high $\fAGN$ galaxies versus low $\fAGN$ galaxies. However, if we had {\it not} corrected $L_IR$ for the AGN contribution when calculating SFE and $\Delta {\rm SFE}_{\rm S14}$, the two-population Kolmogorov-Smirnov test indicates that the distributions of $\Delta {\rm SFE}_{\rm S14}$ for the SF-dominated and AGN-dominated galaxies significantly differ. We show the uncorrected $\Delta {\rm SFE}_{\rm S14}$s as the unfilled symbols in the left panel, and as the purple histogram in the right panel of Figure \ref{SFE}.

The star formation efficiency assumes that star formation is the dominant mode of consuming gas. Powerful AGN are also accreting gas, but at a much lower rate \citep[e.g.][]{delvecchio2014}. Massive outflows could expel gas, but estimates of this mass loss rate exist in only a handful of galaxies. In the local universe, an ionized mass loss rate of $2\,{\rm M_\odot\,yr^{-1}}$ was measured in a Compton thick quasar \citep{revalski2018} and that outflow was solely attributed to its AGN (as opposed to a potential nuclear starburst); such outflow rates are not enough to drastically alter gas depletion timescales in AGN relative to normal SFGs. \citet{gowardhan2018} find massive molecular outflows of hundreds of solar masses per year in two low redshift ($z<0.15$) ultra luminous infrared galaxies. In these galaxies, the molecular outflow is the dominant source depleting the gaseous ISM. However, the outflow correlates with both SFR and AGN luminosity, indicating that the AGN alone is not the dominant predictor of quenching-enabling feedback.

The fact that we do not observe lower SFE in AGN has three possible interpretations. First, our sample may be biased. Considering we have put together a heterogeneous, data-mined sample, we may be missing a key portion of galaxies that would exhibit enhanced or depleted SFE with increasing AGN energetics. However, for the sample that we have, we have taken care to measure properties and identify AGN in a homogeneous way. A large, blind IR survey followed with CO observations for all galaxies is the proper way to test gas depletion timescales in IR AGN.

A second potential reason for the lack of difference between the SFEs may be that IR AGN do not occur at special points in galaxies' life cycles. \citep{kirkpatrick2012} showed that AGN are heating the central ISM relative to normal SFGs (indicated by rising warm dust emission from $\lambda=10-100\mu$m; see also some discussion of AGN heating effects on CO in the following subsection). This heating may prevent star formation in the central 100\,pc, and could be a significant contributor to quenching; verification of this effect requires future resolved observations of the ISM in the centers of IR AGN hosts. However, the lack of evidence for lower gas masses or shorter gas depletion timescales in our AGN suggests that AGN do not play a universal role in quenching star formation, even in massive galaxies \citep[see also][]{xu2015}. 
It leaves open the question of whether AGN are growing in lockstep with their galaxies, consuming the same fuel as the nascent stars \citep[e.\/g.\/,][and references therein]{kormendy2013}. If AGN are passively accreting in their hosts, rather than actively quenching, then why do only some galaxies have visible AGN? Either the triggering mechanisms of IR AGN do not seem to be universally responsible for enhanced star formation in dusty SFGs, as evidenced by our galaxies with high SFRs, short $\tau_{\rm dep}$, but no AGN signatures, or the physics associated with AGN flickering make measuring any correlation hopeless.


Finally, we consider the possibility that the lack of difference between the SFEs may be rooted in the timescales of AGN activity. Theoretical models suggest that supermassive black holes go through growth spurts, causing a variable luminosity on 10,000 year timescales \citep{hickox2014} that would make correlations between AGN activity and SFR (or SFE) difficult to identify. In addition, the density of dust clouds in the torus cause it to cool quickly, from 1000\,K to 100\,K, in less than 10 years, \citep{ichikawa2017}, making IR AGN unidentifiable when the supermassive black hole is not actively accreting. 
For such variable AGN, we may not expect to see a correlation between \fAGN~and SFE. 
The exact timescales traced by AGNs' IR emission is relatively unconstrained, but it is likely extremely short ($<<1$\,Myr). IR-determined SFRs trace timescales of 100 Myr, so we may never see a correlation between the two phenomena, even if there is an underlying relationship. Finally, galaxies evolve over Gyr timescales. Given these differing ranges, looking for correlations between SFE and AGN activity may never reveal any information about the impact of AGN activity on its host galaxy.

Given the limitations potentially introduced by the timescales of AGN activity, we briefly consider whether progress might be made using future far-IR observations. As a back-of-the-envelope example, 
we assume the AGN heating can account for all dust emission at $\lambda\sim70\,\mu$m in a galaxy with $\nu L_\nu=10^{12}\,L_\odot$ at $\lambda=70\,\mu$m. 
If we assume that an AGN behaves as a compact emission source surrounded by a sphere of dust, then we can estimate the radius using $L=4\pi\sigma R^2 T^4$. We use $T=100\,$K, which gives $R\sim100$\,pc. This is similar to the radius of the narrow line region, which extends 10-1000\,pc \citep{groves2006,hainline2013}. 
At the extreme, if the AGN is heating the local diffuse cold neutral medium, beyond the ionized NLR, the main cooling is through the 
[CII] fine structure line. 
For thermal equilibrium, with a low density of
$n=20\,{\rm cm}^{-3}$, the cooling timescale is longer than 10,000 years. However, if the AGN's influence is confined to the ionized narrow line region, then the cooling timescale will be much shorter \citep{groves2006,hainline2016}. Therefore, resolved 70\,$\mu$m observations (on scales of a kpc or less) may be a promising way to look for AGN emission over long timescales. 

\subsection{Average Gas Excitation Properties of AGN- and Starbust-dominated Galaxies}
\label{sec:agn_sled}

\begin{figure}
\epsscale{1.0}
\plotone{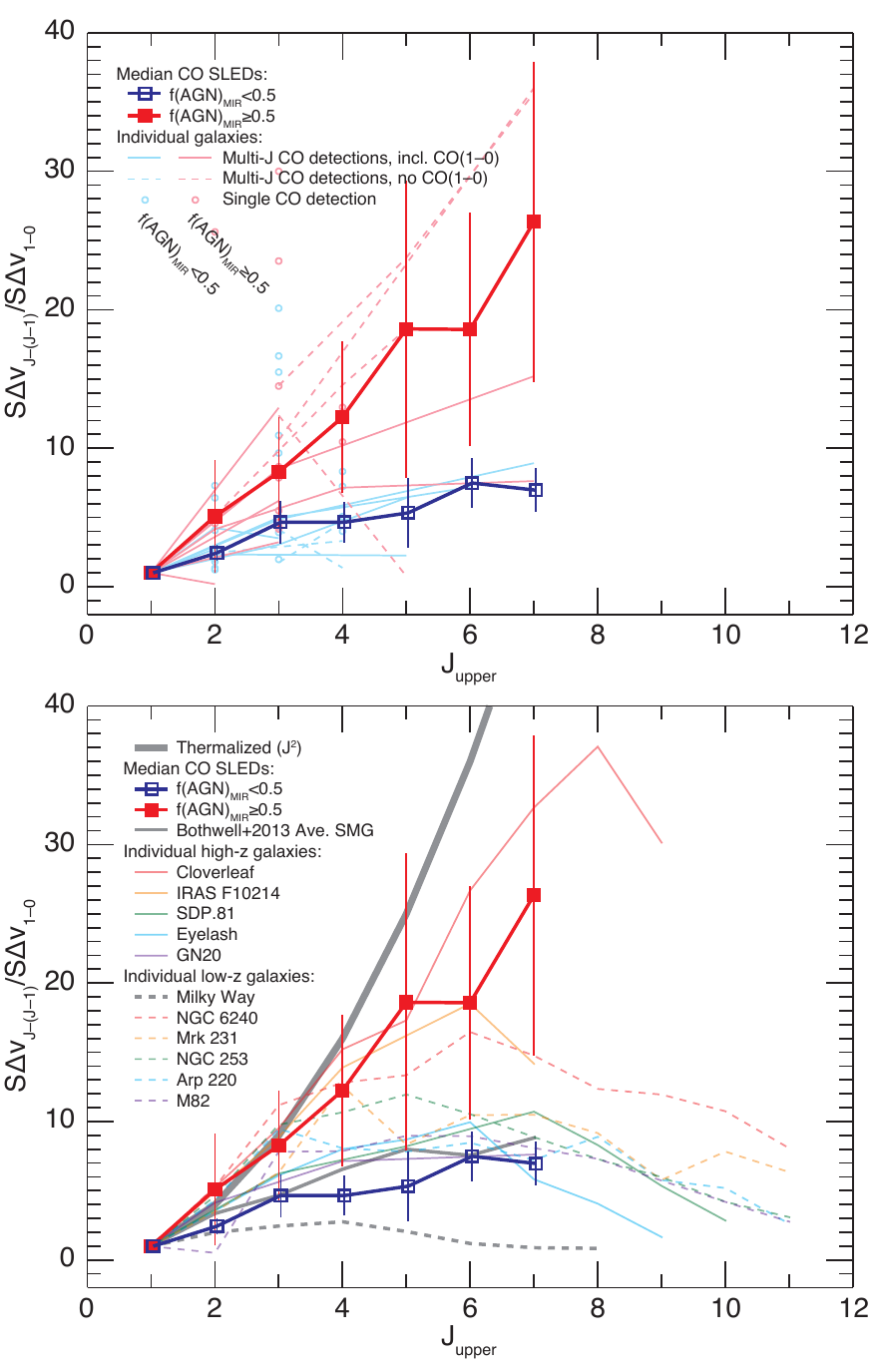}
\caption{Median CO SLEDs and line ratio uncertainties for sources with mid-IR AGN fractions $\ge50\%$ (red filled squares/thick lines) and mid-IR AGN fractions $<50\%$ (dark blue open squares/thick lines). \emph{Top:} The CO detections for individual galaxies contributing each median CO SLED are shown in light red circles/thin lines for the $\ge50\%$ mid-IR AGN fractions and light blue circles/thin lines for the $<50\%$ mid-IR AGN fractions. Sources with only a single reported CO detection are shown as light color circles; the SLED of sources that lack published \mbox{CO(1--0)} measurements are shown with dashed lines (both scaled by the corresponding sub-samples' median \mbox{CO(1--0)} fluxes as described in Section~\ref{gas}). \emph{Bottom:} We compare the same median CO SLEDs in the top panel to the SLEDs of several well-known high-redshift galaxies (solid colored thin lines) and low-redshift galaxies (dashed colored thin lines) as listed in the legend. CO SLEDs for the high-redshift galaxies are taken from the samples described in Section~\ref{excitation} that are used to calculate gas mass excitation corrections. CO SLEDs for low-redshift sources are taken from \citet{kamenetzky2016}, except for the Milky Way (dashed gray line) which is from \citet{fixsen1999}. We also show the average SLED for SMGs calculated in \citet{bothwell2013} (solid gray line), and the line ratios for thermalized emission (thick gray line). \label{fig:fmirsleds}}
\end{figure}

Using the methods described in Section~\ref{gas}, we also calculate the median CO SLEDs for galaxies with measured MIR AGN fractions (Table~\ref{COratios2} and Figure~\ref{fig:fmirsleds}). While the effects of AGN heating are likely continuous with \fAGN, there are too few CO detection per source to look for fine-grained differences in the CO SLEDs as a function of \fAGN. Therefore, as in the previous section, we divide the sources into two groups at $\fAGN=0.5$ in order to improve our statistics; for convenience, we refer to the sources below and above $\fAGN=0.5$ as SF-dominated and AGN-dominated respectively.
In the bottom panel of Figure~\ref{fig:fmirsleds}, we compare our new SLEDs to previous work (SLEDs for local ULIRGs are taken from \citealt{kamenetzky2016}). Our median SF-dominated SLED closely resembles the average SMG SLED from \citet{bothwell2013} and is also consistent with the SLEDs of local U/LIRGs such as Arp 220 and M82 (given our large uncertainties and the effects of small number statistics). Our median AGN-dominated SLED appears consistent with thermalized emission in the Rayleigh-Jeans limit (where flux ratios are equal to $J^2$). While thermalized emission is expected to arise in the dense molecular ISM of ULIRGs, SMGs, and AGN for low-$J$ lines \citep[e.\/g.\/,][]{weiss2007a,harris2010,riechers2011f}, it is generally not expected for the kinetic temperature to be sufficiently high to drive higher-$J$ emission to the asymptotic Rayleigh-Jeans limit. The consistency with thermalized emission for the high-$J$ lines in the AGN-dominated sample is likely due to the small number statistics that produce the considerable uncertainties. Significant amounts of high excitation emission is not uncommon for high-$z$ AGN, and our AGN-dominated SLED is consistent with that of the Cloverleaf galaxy and IRAS F10214+4724, which are both type 1 QSOs at $z>2$. Local U/LIRGs with known AGN (i.e. NGC 6240, Mrk 231, and NGC 253) appear to have somewhat lower line ratios in their highest-$J$ transitions than we find in our median AGN-dominated SLED, but again they are consistent given our sample's small numbers and large uncertainties.  Given the uncertainties for our line ratios (Table~\ref{COratios2}), there is no significant difference between the CO excitation for the AGN-dominated galaxies ($\fAGN\geq0.5$) and the SF-dominated galaxies ($\fAGN<0.5$).


\begin{deluxetable}{@{\extracolsep{4pt}} ccccc}
\tablecaption{Median line ratios for AGN and star formation dominated galaxies \label{COratios2}}
\tablehead{ & \multicolumn{2}{c}{$f(AGN)_{MIR}<50\%$} & \multicolumn{2}{c}{$f(AGN)_{MIR}\geq50\%$} \\
\cline{2-3} \cline{4-5}
\colhead{Transition} & \colhead{Ratio} & \colhead{$N$} & \colhead{Ratio} & \colhead{$N$}}
\startdata
CO(1--0) & \nodata & 8 & \nodata & 8 \\
$r_{2,1}$ & $0.61\pm0.15$ & 16 & $1.27\pm1.01$ & 5 \\ 
$r_{3,1}$ & $0.52\pm0.17$ & 15 & $0.92\pm0.44$ & 18 \\  
$r_{4,1}$ & $0.29\pm0.09$ & 7 & $0.76\pm0.34$ & 5 \\ 
$r_{5,1}$ & $0.21\pm0.10$ & 2 & $0.74\pm0.43$ & 3 \\ 
$r_{6,1}$ & $0.21\pm0.05$ & 1 & $0.52\pm0.23$ & 1 \\ 
$r_{7,1}$ & $0.14\pm0.03$ & 1 & $0.54\pm0.24$ & 4
\enddata
\end{deluxetable}

\begin{figure}
\epsscale{1.0}
\plotone{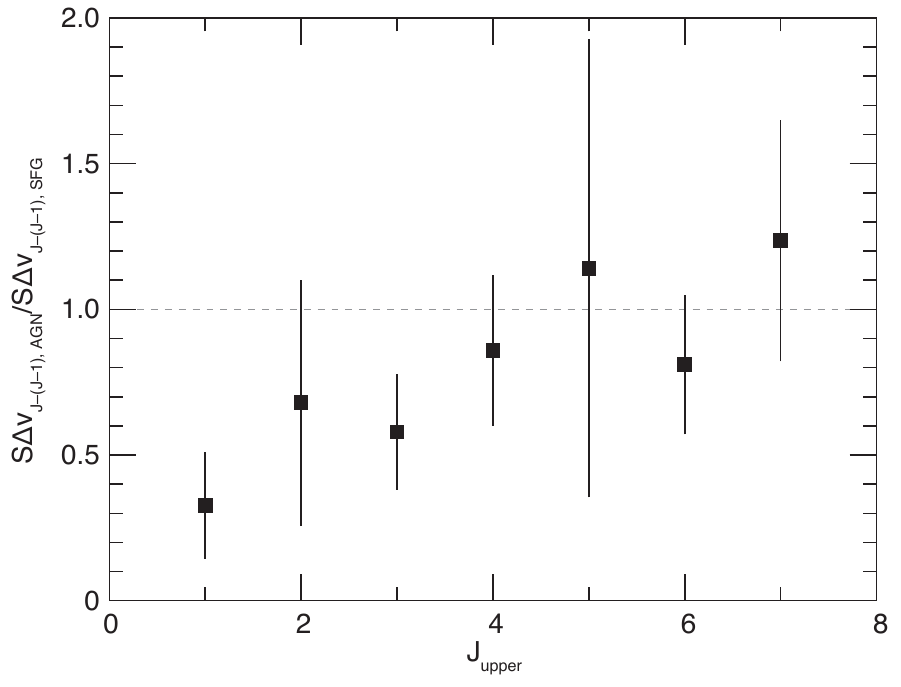}
\caption{Median CO line fluxes for the galaxies with mid-IR AGN fractions $\ge50\%$ in ratio to the median CO line fluxes for the galaxies with mid-IR AGN fractions $<50\%$. For clarity we show a gray dashed line for equal line fluxes between the two populations. \label{fig:agn2sfratios}}
\end{figure}

However, there may be a systematic difference in the CO SLED for the two populations (Figure~\ref{fig:fmirsleds}) even if individual line ratios are consistent within their uncertainty. In order to explore whether this systematic difference is significant, we compare the average emission line flux between the two populations for each CO line directly, thereby avoiding additional uncertainty from the \mbox{CO(1--0)} line when comparing the ratios between the two populations. We use a similar bootstrapping with replacement technique as described in Section~\ref{gas} to determine the median line luminosities and their uncertainties for both samples, where line fluxes have been scaled linearly by the ratio of their individual IR luminosity to the median IR luminosity of the full sample (Figure~\ref{fig:agn2sfratios}). While there are still substantial uncertainties on the population ratios for each line, the observed trend with $J$ is statistically significant; performing a Spearman's rank correlation test yields a $1.4\%$ chance of getting an observed correlation at least this strong with a null hypothesis of no correlation. However, this correlation is not robust. When removing either of the extrema from the correlation test (either the \mbox{CO(1--0)} population ratio or the \mbox{CO(7--6)} population ratio), the probability increases to $7.6\%$, which is sufficiently large that we likely cannot rule out the null hypothesis of no correlation. In addition, if we Monte Carlo over the uncertainties (since the Spearman's rank test does not factor those in), only $27\%$ of iterations yield $p<0.05$ of getting an observed correlation at least this strong. Therefore we must conclude that our data does not show a significant difference between the CO excitation of SF-dominated and AGN-dominated galaxies as defined by $\fAGN=0.5$. 

Ultimately, only two lines have $>5$ detections in both the AGN-dominated and SF-dominated samples: the \mbox{CO(1--0)} line with eight detections below and above $\fAGN=0.6$, and the \mbox{CO(3--2)} line with 15 and 18 detections below and above $\fAGN=0.6$ respectively (Table~\ref{COratios2}). We find that the ratio of \mbox{CO(1--0)} fluxes for the AGN-dominated to SF-dominated galaxies is $0.33\pm0.18$, and the ratio for the \mbox{CO(3--2)} line is $0.58\pm0.20$. At face value, these low-J ratios seem to indicate that for the same IR luminosity, $z>1$ galaxies with SF-dominated MIR fluxes have more gas than $z>1$ galaxies with AGN-dominated MIR fluxes. Such systematically lower gas masses in high-$z$ AGN might be expected if AGN play a role in quenching periods of rapid star formation or if AGN are temporally correlated with the transition between more bursty and more quiescent phases of galaxy evolution. However, this result would be at odds with the observed lack of difference in SFEs found in Section~\ref{sec:sfe}. Due to the small sample size, more data is needed before drawing strong conclusions.

This observed difference may be artificially produced by our method of linearly scaling  $L^\prime_{\rm CO}$ using the ratio of the total $L_{\rm IR}$ (including both the contributions from star formation and any AGN) to some fiducial value.\footnote{Scaling by the $L_{\rm IR}^{\rm SF}$ do not change these result.} If the correlation between $L_{\rm IR}$ and $L^\prime_{\rm CO}$ (the integrated Schmidt-Kennicutt relation) is super-linear (or sub-linear in the case of Fig.~\ref{lir_lco} where the axes are swapped), and $L_{\rm IR}$ of the AGN-dominated galaxies are $\sim0.5-1$ dex brighter than the SF-dominated galaxies on average, then in the process of linearly scaling $L^\prime_{\rm CO}$ we effectively over-scale the gas emission of the AGN-dominated sources and correspondingly under-scale that for the SF-dominated sources. These scalings would then compound to produce the relatively lower ratios of AGN-to-SF gas masses. This effect of the scaling may come into play in the average SLEDs produced in Section~\ref{gas}, since AGN-dominated and SF-dominated galaxies are averaged together. If the total $L_{\rm IR}$ distributions differ systematically with galaxy class (or some other parameter of interest) that has systematically differing CO excitation, and the $L_{\rm IR}$-$L^\prime_{\rm CO}$ is super-linear, then the net effect would be a bias towards the excitation characteristics of the lower luminosity population. 
If the index of the Schmidt-Kennicutt correlation becomes shallower for higher-$J$ CO gas tracers (e.\/g.\/, \citealt{greve2014} and references therein; cf. \citealt{liu2015,kamenetzky2016}), then the applied scaling becomes systematically more appropriate for the higher-$J$ CO lines, and thus could also produce the observed correlation in Figure~\ref{fig:agn2sfratios}. In order to correct for such $L_{\rm IR}$-dependent effects, either we would need new observations of a $L_{\rm IR}$-matched sample with MIR-determined AGN fractions, or we would need to find some other parameter that linearly scales with gas mass to use when calculating the median CO luminosities. However, such scaling effects produced by the different populations' $L_{\rm IR}$ distributions are removed by considering the higher-$J$ lines in ratio to the ground state (as for the CO SLED plots in Figure~\ref{fig:fmirsleds}). 

While high-$J$ CO emission from IR-bright AGN is well-documented \citep[e.\/g.\/,][]{downes1999,bertoldi2003,walter2003,weiss2007b,gallerani2014,tuan-anh2017}, we do not find conclusive evidence for systematically higher excitation CO SLEDs in this AGN-dominated \fAGN\ sample. In spatially resolved studies of galaxies, it is well known that regions with different heating sources (and thus cloud physical conditions; i.\/e. temperatures and densities) will produce different CO line ratios---cold molecular clouds with moderate star formation rates in the peripheries of disk galaxies will produce sub-thermal emission at even the lowest-$J$ CO lines (e.\/g.\/, $r_{3,1}=0.4$ for the outer disk of the Milky Way; \citealt{fixsen1999}) while dense starbursts' clouds or the central regions near AGN can produce nearly thermalized emission to much higher-$J$ (e.\/g.\/, $r_{3,1}$ variations in the Antennae or near the central AGN of NGC 1068; \citealt{zhu2003,spinoglio2012b,viti2014}). Therefore, depending on the relative balance of these energy sources in each galaxy, we expect to measure different global line ratios \citep[e.\/g.\/,][]{harris2010,narayanan2014}. It follows that if the warm dust continuum emission from the AGN is dominating $L_{\rm IR}$, and dust and molecular gas is comingled, then we might expect to see higher excitation is such systems' CO SLEDs. Given the limitations of our current sample, we do not see evidence of this effect when comparing high-$z$ SFGs and IR AGN as defined by \fAGN. Therefore, barring access to an $L_{\rm IR}$-matched sample with MIR-determined AGN fractions (which will become possible in the era of JWST), increasing the number of multi-line CO detections (particularly at higher-$J$) for archival {\it Spitzer} sources is likely the best way to improve statistics for the immediate future.

\subsection{Evolving Gas-to-Dust Ratio}
\label{sec:gdr}

Due to the relatively modest time required for continuum observations as compared with gas emission lines submillimeter emission has recently been promoted as a gas mass tracer \citep[e.\/g.\/,][]{groves2015,scoville2016}. Calibrators for converting dust continuum emission into a gas mass rely on the assumption that the gas-to-dust mass ratio (GDR) is constant in solar metallicity galaxies out to $z\sim4$. We can test this assumption, since we have independently measured dust and gas masses in a heterogeneous sample of galaxies out to $z\approx4$, and we can look for changes in the GDR with increasing \fAGN.  

\begin{figure}
    \centering
    \includegraphics[width=3.3in]{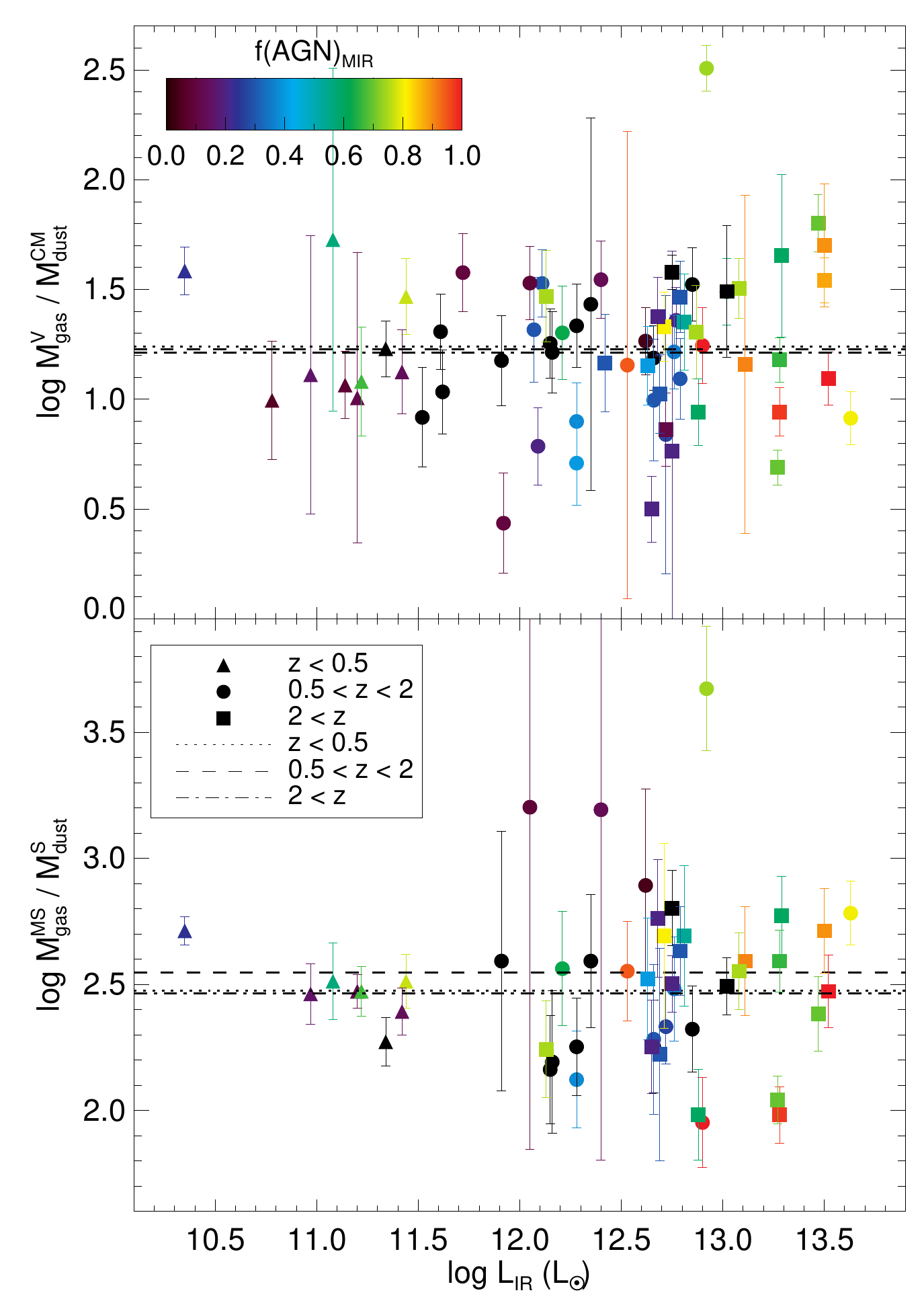}
    \caption{Gas to dust ratio (GDR) calculated with two different $M_{\rm gas}$ and $M_{\rm dust}$. Regardless of the methodology, GDR is consistent across redshift, $L_{\rm IR}$, and AGN presence. {\it Top--}$M_{\rm gas}^{\rm V}$ is calculated with a variable $\alpha_{CO}$ that depends on metallicity and ISM surface density. $M_{\rm dust}^{CM}$ is calculated by fitting far-IR observations with a dust radiative transfer model. We calculate the mean GDR values for three redshift bins: $z<0.5$ (dotted line), $0.5\geq z<2$ (dashed line), $2\geq z$ (dot-dashed line). The mean values are all lower than expected ($\sim20$). {\it Bottom--}GDR when $\alpha_{CO}=4.6$ is used. $M_{\rm dust}^{S}$ is calculated from submm flux assuming $T_C=25\,$K. The mean GDR for all three redshift bins is remarkably consistent $\sim300$. }
    \label{gdr}
\end{figure}

Figure \ref{gdr} shows the GDR for our sample as a function of $L_{\rm IR}$. In the top panel, we use a varying $\alpha_{\rm CO}$ \citep{narayanan2012}, and in the bottom panel, we use a constant $\alpha_{\rm CO}=4.6\,\rm{M_\sun\,(K\,km\,s^{-1}\,pc^2 )^{-1}}$ for all galaxies, as described in Section \ref{sec:alpha}. We determine the geometric mean of the GDR in three redshift bins and find it to be remarkably consistent; for a varying (constant) CO-to-${\rm H_2}$ conversion factor, the GDR is $17\pm3$ ($300\pm28$) for $z<0.5$, $17\pm3$ ($353\pm81$) for $0.5\leq z<2$, and $16\pm3$ ($292\pm34$) for $z\geq2$. We find no trend in GDR with $L_{\rm IR}$, and no difference in GDR between the sources above and below $\fAGN=0.5$. The exact gas to dust ratio depends strongly on the assumed $\alpha_{\rm CO}$. It is difficult to know what the expected GDR at $z\sim1-3$ is, since it hinges so sensitively on $\alpha_{\rm CO}$ and converting higher CO transitions to the ground state. In a sample of four solar metallicity galaxies at $z\sim1.4$, \citet{seko2016} use the CO(2-1) transition and $\alpha_{CO}=4.36$ and calculate GDRs in the range 220-1450, in agreement with our measurements. \citet{saintonge2013} measured the GDR in 17 lensed galaxies at $z=1.4-3.1$. They observe CO(3-2) and use a variable $\alpha_{\rm CO}$ based on metallicity. They derive gas to dust ratios in the range of 100-700 and demonstrate that these values are 1.7$\times$ greater than local galaxies. For local galaxies, \citet{leroy2011} presents a relation to derive GDR from metallicity. Based on the \cite{leroy2011} relation, the expected average GDR of our galaxies is $50$, closer to the value we derived following the \citet{narayanan2012} method. The expected GDR, based on literature values for solar metallicity galaxies, is 100--200. The use of $M_{\rm dust}^{\rm CM}$ rather than $M_{\rm dust}^{\rm S}$ would increase the GDR by a small amount (the most significant increase is for $0.5\leq z<2$ where the GDR increases to $\sim350$).

The GDR calculted using the varying $\alpha_{\rm CO}$ are very low. These low GDRs are due to the low gas masses inferred from the lower CO-to-${\rm H_2}$ conversion factors; these low conversion factors are either due to small radii used in calculating the CO surface brightness ($\Sigma_{\rm CO}$) or overestimated metallicities. Large $\Sigma_{\rm CO}$ (and thus low GDRs) could occur if dust and gas is not cospatial and the dust emission in high-$z$ SMGs is particularly compact \citep[e.\/g.\/,][]{hodge2016}. If we assume that all galaxies have solar metallicity, then the average radius of the CO emission would need to be a factor of ten larger than the typical dust emission to produce GDR$\sim100-200$ in all galaxies. Indeed, if we assume that all galaxies have solar metallicity, then there would be a consistent increase in GDR with redshift, so that by $z\geq 2$, GDR=34. Alternatively, if the low GDR is due to an overestimate of the metallicities, we would need to drop metallicities to $Z<8.0$ in order to produce GDRs of $\sim100$. Such a drop in metallicity would imply a significant evolution in the relationship between $M_{\rm dust}$ and metallicity in massive galaxies as a function of redshift. 

Assuming that any concerns with the CO-to-${\rm H_2}$ conversion factor are satisfied, our values of the GDR indicate that the varying assumptions used for calculating $M_{\rm gas}$ and $M_{\rm dust}$ likely require adjustment for galaxies beyond the local universe. Additional concerns about $\alpha_{\rm CO}$
arise from estimating the gas phases at high redshift, particularly since more hydrogen may be in the molecular phase rather than the atomic phase compared with the local universe \citep{bertemes2018,janowiecki2018}. Regardless of method or assumptions, the GDR is consistent for all \fAGN\ across the  range of redshifts probed here. However, for individual objects, the ratios can vary significantly. Using the dust emission to estimate gas emission is likely only valid on average across large samples, rather than for individual objects.

\section{Conclusions \& Future Directions}
\label{conclusions}

We have compiled a heterogeneous sample of 67 galaxies from the literature that have CO line measurements and IR spectroscopy/photometry from $z=0-4$.
We have diagnosed the presence of an IR AGN using spectral decomposition or IR color selection based on techniques calibrated for galaxies at $z\sim1-3$, and measured $L_{\rm IR}$, $M_{\rm dust}$, and $L^\prime_{\rm CO}$ consistently for all sources. Our findings are as follows:
\begin{enumerate}
    \item SFE is not strongly correlated with AGN emission. Strong AGN are observed to have both high and low star formation efficiencies, indicating that the presence of an AGN itself does not indicate a galaxy is beginning to quench. Any expected correlation may necessarily be murky due to AGN luminosity flickering on much shorter timescales than galaxy evolution. 
    \item It is important to fully sample the IR SED, use appropriate templates, and take AGN heating effects into account when calculating $L_{\rm IR}$. If this is not done, derived SFRs will be too high and SFE will wrongly appear to be enhanced in IR AGN. 
    \item Gas-to-dust ratios do not appear to evolve with redshift for this heterogeneous sample, although the scatter is large. We find consistent gas-to-dust ratios regardless of redshift, luminosity, or AGN strength. However, for individual objects, the exact gas to dust ratio can vary significantly.
    \item We do not find a statistically significant difference in the CO excitation for individual line ratios for galaxies above and below $\fAGN=0.5$. While it appears that the median CO SLED of sources with $\fAGN\geq0.5$ is at systematically higher excitation than median CO SLED of galaxies with $\fAGN<0.5$ for all rotational transitions, this result is not robust statistically.
\end{enumerate}

Many of our conclusions about the molecular gas content and the effects of AGN heating is limited by small number statistics and the lack of matched luminosity samples. A systematic CO SLED survey of galaxies with known IR AGN content (i.\/e.\ measured \fAGN) is critical for disentangling evidence of AGN feedback affecting star-forming molecular gas via the gas physical conditions. Since the problem of high SFRs that require some limiting mechanism to prevent the over-production of massive galaxies today is particularly acute among dusty SFGs, studies of AGN feedback are more likely to be successful when focused on high-$z$ IR-bright galaxies such as SMGs. Since comparing average gas properties requires scaling out the dominant effect of brighter/more-massive galaxies being brighter across the electromagnetic spectrum (to first order), accurate normalization is similarly critical. Firmer conclusions could be reached with either (a) more observations of the \mbox{CO(1--0)} line to normalize the excitation, (b) more observations of the $J\geq5$ CO lines where the number of detections for \fAGN~samples are lowest, or (c) developing matched IR luminosity samples among IR AGN and SFGs. Until we have the mid-IR spectroscopic/photometric capabilities of JWST that enable the selection of matched $L_{\rm IR}$ samples of high-$z$ galaxies, we are restricted to analyzing galaxies with archival {\it Sptizer} data.

\vspace{0.5cm}
AK thanks Sukanya Chakrabarti and Desika Narayanan for helpful conversations. AK gratefully acknowledges support from the YCAA Prize Postdoctoral Fellowship.

\appendix 

\begin{deluxetable}{lllcccc}
\tablecolumns{7}
\tablecaption{Sample Positions\label{properties}}
\tablehead{\colhead{Name} & \colhead{RA} & \colhead{Dec} & \colhead{\it z} & \colhead{\fAGN\tablenotemark{a}} & \colhead{$\log L_{\rm IR} (L_\odot)$} & 
\colhead{$L^\prime_{\rm CO}\,({\rm K\,km\,s^{-1}\,pc^2})$} }
\startdata
5MUSES-179 & 16:08:03.71 & +54:53:02.0 & 0.053 & 0.24\tablenotemark{a} & $10.35\pm0.03$ & \,\ $9.16\pm0.04$ \\
5MUSES-169 & 16:04:08.30 & +54:58:13.1 & 0.064 & 0.06\tablenotemark{a} & $10.78\pm0.18$ & \,\ $8.92\pm0.10$ \\
5MUSES-105 & 10:44:32.94 & +56:40:41.6 & 0.068 & 0.16\tablenotemark{a} & $10.97\pm0.05$ & \,\ $9.19\pm0.08$ \\
5MUSES-171 & 16:04:40.64 & +55:34:09.3 & 0.078 & 0.08\tablenotemark{a} & $11.14\pm0.14$ & \,\ $9.33\pm0.08$ \\
5MUSES-229 & 16:18:19.31 & +54:18:59.1 & 0.082 & 0.12\tablenotemark{a} & $11.20\pm0.04$ & \,\ $9.55\pm0.06$\\
5MUSES-132 & 10:52:06.56 & +58:09:47.1 & 0.117 & 0.00\tablenotemark{a} & $11.34\pm0.05$ & \,\ $9.44\pm0.09$\\
5MUSES-227 & 16:17:59.22 & +54:15:01.3 & 0.134 & 0.57\tablenotemark{a} & $11.08\pm0.05$ & \,\ $9.00\pm0.13$\\
5MUSES-141 & 10:57:05.43 & +58:04:37.4 & 0.140 & 0.67\tablenotemark{a} & $11.22\pm0.01$ & \,\ $9.46\pm0.09$ \\
5MUSES-136 & 10:54:21.65 & +58:23:44.7 & 0.204 & 0.77\tablenotemark{a} & $11.44\pm0.04$ & \,\ $9.67\pm0.10$\\
5MUSES-216 & 16:15:51.45 & +54:15:36.0 & 0.215 & 0.14\tablenotemark{a} & $11.42\pm0.05$ & \,\ $9.62\pm0.09$\\
PEPJ123646+621141        & 12:36:46.18 & +62:11:42.0 & 1.016 & 0.00\,\,\, & $11.62\pm0.11$ & \,\ $9.91\pm0.15$\\
GN70.38\tablenotemark{a} & 12:36:33.68 & +62:10:05.9 & 1.016 & 0.00\tablenotemark{a} & $11.91\pm0.08$ & $10.20\pm0.16$  \\
EGS13035123				 & 14:20:05.40 & +53:01:15.4 & 1.070 & 0.20\,\,\,            & $12.09\pm0.08$ & $10.39\pm0.12$ \\
PEPJ123759+621732		 & 12:37:59.47 & +62:17:32.9 & 1.084 & 0.00\,\,\,            & $11.52\pm0.30$ & $10.01\pm0.16$\\
GN70.8                  & 12:36:20.94 &	+62:07:14.2 & 1.148 & 0.00\tablenotemark{a}  & $12.16\pm0.01$ & $10.17\pm0.15$ \\
PEPJ123750+621600       & 12:37:50.89 & +62:16:00.7 & 1.170 & 0.00\,\,\, & $11.61\pm0.22$ & $10.16\pm0.13$ \\
SMMJ163658+405728		& 16:36:58.78 & +40:57:27.6 & 1.193 & 0.38\tablenotemark{a} & $12.28\pm0.20$ & $10.35\pm0.15$ \\
EGS13003805				& 14:19:40.08 & +52:49:38.6 & 1.200 & 0.30\,\,\, & $12.11\pm0.20$ & $10.54\pm0.13$ \\
J123634.53+621241.3     & 12:36:34.57 & +62:12:41.0 & 1.223 & 0.04\tablenotemark{a} & $12.62\pm0.01$ & $10.69\pm0.13$ \\
GN70.104                & 12:37:02.74 & +63:14:01.5 & 1.246 & 0.00\tablenotemark{a} & $12.15\pm0.08$ & $10.24\pm0.13$ \\
PEPJ123712+621753       & 12:37:12.15 & +62:17:54.0 & 1.249 & 0.10\,\,\, & $11.72\pm0.10$ & \,\ $9.66\pm0.16$\\
SMMJ030227+000653       & 03:02:27.66 & +00:06:53.0 & 1.406 & 0.00\tablenotemark{a} & $12.66\pm0.08$ & $10.43\pm0.13$ \\
EGS13004291				& 14:19:14.96 & +52:49:30.1 & 1.410 & 0.20\,\,\, & $12.77\pm0.03$ & $10.73\pm0.12$ \\
no.226                  & 03:32:15.99 & -27:48:59.4 & 1.413 & 0.30\,\,\, & $12.07\pm0.10$ & $10.05\pm0.20$ \\
RGJ123645.88+620754.2   & 12:36:45.88 & +62:07:54.2 & 1.434 & 0.13\tablenotemark{a} & $12.40\pm0.01$ & $10.41\pm0.16$ \\
3C298                   & 14:19:08.18 & +06:28:34.8 & 1.438 & 0.80\,\,\, & $13.63\pm0.07$ & $11.63\pm0.12$ \\
HR10                    & 16:45:02.26 &	+46:26:26.5 & 1.440 & 0.27\tablenotemark{a} & $12.72\pm0.09$ & $10.82\pm0.07$ \\
BzK-4171                & 12:36:26.52 & +62:08:35.4 & 1.465 & 0.09\tablenotemark{a} & $12.05\pm0.09$ & $10.40\pm0.14$ \\
51613					& 10:02:43.36 & +01:34:20.9 & 1.517 & 0.10\,\,\, & $11.92\pm0.10$ & $10.38\pm0.11$ \\
BzK-21000			    & 12:37:10.60 & +62:22:34.6 & 1.523 & 0.00\tablenotemark{a} & $12.35\pm0.08$ & $10.39\pm0.11$ \\
51858					& 10:02:40.43 & +01:34:13.1 & 1.556 & 0.40\,\,\, & $12.28\pm0.10$ & $10.10\pm0.13$ \\
MIPS8342                & 17:14:11.55 & +60:11:09.3 & 1.562 & \ \ 0.74\tablenotemark{a,b} & $12.92\pm0.05$ & $11.66\pm0.09$\\ 
3C318                   & 05:20:05.49 &	+20:16:05.7 & 1.577 & 1.00\tablenotemark{a} & $12.90\pm0.10$ & $10.76\pm0.14$\\
SMMJ105151+572636       & 10:51:51.77 & +57:26:35.3 & 1.597 & 0.30\tablenotemark{a} & $12.66\pm0.08$ & $10.47\pm0.25$ \\
COSB011                 & 10:00:38.02 & +02:08:22.9 & 1.827 & 0.60\,\,\, & $13.02\pm0.17$ & $10.92\pm0.14$ \\
SMMJ123555+620901       & 12:35:54.85 & +62:08:54.7 & 1.864 & \ \ 0.63\tablenotemark{a,b} & $12.21\pm0.12$ & $10.82\pm0.17$ \\
SMMJ123632+620800		& 12:36:33.04 & +62:08:05.1 & 1.994 & \ \ 0.95\tablenotemark{a,b} & $12.53\pm0.12$ & $10.81\pm0.17$ \\
SMMJ123712+621322       & 12:37:12.12 & +62:13:22.2 & 1.996 & 0.00\tablenotemark{a} & $12.28\pm0.08$ & $10.66\pm0.18$\\
SMMJ123618.33+621550.5  & 12:36:18.47 & +62:15:51.0 & 1.996 & 0.36\tablenotemark{a} & $12.76\pm0.01$ & $10.70\pm0.16$ \\
SMMJ123711+621325       & 12:37:11.19 & +62:13:31.2 & 1.996 & 0.00\tablenotemark{a} & $12.85\pm0.01$ & $10.86\pm0.16$\\
RG J123711              & 12:37:11.34 & +62:32:31.0 & 1.996 & 0.30\,\,\, & $12.79\pm0.20$ & $10.43\pm0.18$\\
MIPS16080               & 17:18:44.77 & +60:01:15.9 & 2.006 & \ \ 0.81\tablenotemark{a,b} & $12.71\pm0.02$ & $10.51\pm0.14$ \\
SMMJ021738-050339       & 02:17:38.68 & -05:03:39.3 & 2.037 & 0.20\,\,\, & $12.68\pm0.05$ & $10.94\pm0.16$ \\
RG J123644.13+621450.7  & 12:36:44.13 & +62:14:50.7 & 2.095 & 0.75\tablenotemark{a} & $12.13\pm0.02$ & $10.52\pm0.17$ \\
HS1002+4400				& 10:05:17.43 & +43:46:09.3 & 2.102 & 1.00\tablenotemark{a} & $13.52\pm0.10$ & $11.07\pm0.12$ \\
MIPS15949               & 17:21:09.22 & +60:15:01.3 & 2.119 & \ \ 0.91\tablenotemark{a,b} & $13.11\pm0.02$ & $10.66\pm0.12$\\
MIPS16144			 	& 17:24:22.09 & +59:31:50.8 & 2.128 & 0.11\tablenotemark{a} & $12.72\pm0.12$ & $10.79\pm0.14$ \\
MIPS429                 & 17:16:11.83 & +59:12:13.2 & 2.201 & 0.54\tablenotemark{a} & $12.81\pm0.05$ & $10.66\pm0.14$ \\
SMMJ123549+6215         & 12:35:49.43 & +62:15:36.7 & 2.202 & 0.20\,\,\, & $12.75\pm0.05$ & $10.88\pm0.05$ \\
RXJ124913.86-055906.2   & 12:49:13.85 & -05:59:19.4 & 2.247 & 1.00\tablenotemark{a} & $13.95\pm0.09$ & $10.72\pm0.10$ \\
SMMJ021725-045934       & 02:17:25.16 & -04:59:34.9 & 2.292 & 0.40\,\,\, & $12.63\pm0.06$ & $10.68\pm0.16$  \\
MIPS16059				& 17:24:28.45 & +60:15:33.0 & 2.326 & \ \ 0.75\tablenotemark{a,b} & $12.87\pm0.15$ & $10.45\pm0.14$ \\
SMMJ16371+4053          & 16:37:06.50 & +40:53:14.0 & 2.377 & 0.60\tablenotemark{b} & $12.88\pm0.07$ & $10.53\pm0.15$ \\
SMMJ163650+405734       & 16:36:50.41 & +40:57:34.4 & 2.385 & \ \ 0.75\tablenotemark{a,b} & $13.08\pm0.11$ & $11.01\pm0.13$\\
SMMJ105227+572512       & 10:52:27.10 & +57:25:16.0 & 2.440 & 0.30\,\,\, & $12.69\pm0.06$ & $10.38\pm0.16$ \\
SMMJ16358+4105          & 16:36:58.20 & +41:05:24.0 & 2.450 & 0.00\,\,\, & $13.02\pm0.06$ & $11.03\pm0.08$ \\
SMMJ123707+6214         & 12:37:07.28 &	+62:14:08.6 & 2.488 & 0.00\tablenotemark{a} & $12.75\pm0.08$ & $10.99\pm0.05$ \\
SMMJ123606+621047       & 12:36:06.85 & +62:10:21.4 & 2.505 & 0.30\,\,\, & $12.42\pm0.13$ & $10.41\pm0.18$ \\
SMMJ221804+002154       & 22:18:04.42 &	+00:21:54.4 & 2.517 & 0.20\,\,\, & $12.65\pm0.10$ & $10.71\pm0.15$ \\
SMMJ021738-050528       & 02:17:38.88 & -05:05:28.2 & 2.541 & 0.30\,\,\, & $12.79\pm0.01$ & $10.99\pm0.15$ \\
VCV J1409+5628			& 14:09:55.50 &	+56:28:27.0 & 2.583 & 0.98\tablenotemark{a} & $13.28\pm0.09$ & $10.93\pm0.11$\\
AMS12                   & 17:18:22.65 & +59:01:54.3 & 2.766  & 0.70\,\,\, & $13.47\pm0.04$ & $10.87\pm0.13$  \\
SMMJ04135+10277         & 04:13:27.28 & +10:27:40.4 & 2.846 & 0.70\tablenotemark{b} & $13.27\pm0.02$ & $10.94\pm0.08$ \\
B3 J2330+3927           & 23:30:24.84 & +39:37:12.2 & 3.094 & 0.90\,\,\, & $13.50\pm0.03$ & $11.38\pm0.16$ \\
6C1909+722				& 19:08:23.70 &	+72:20:11.6 & 3.532 & 0.88\tablenotemark{a} & $13.50\pm0.12$ & $11.07\pm0.10$ \\
4C41.17                 & 06:50:52.10 & +41:30:30.5 & 3.796 & 0.60\,\,\, & $13.29\pm0.06$ & $11.26\pm0.15$ \\
GN20                    & 12:37:11.86 & +62:22:12.6 & 4.055 & 0.66\tablenotemark{a} & $13.28\pm0.09$ & $11.15\pm0.10$
\enddata
\tablenotetext{a}{These sources had {\it Spitzer} IRS spectroscopy which was decomposed to measure the mid-IR AGN fraction.}
\tablenotetext{b}{We identified these sources as IR AGN, but they are classified as SFGs in \citet{carilli2013}.}
\end{deluxetable}

\begin{deluxetable}{lcccccc}
\tablecolumns{7}
\tablecaption{Sample Derived Gas and Dust Properties\label{dat_table}}
\tablehead{\colhead{Name} & \colhead{$\alpha_{\rm CO}$\tablenotemark{a}} & \colhead{$R$ (kpc)\tablenotemark{b}} & \colhead{$\log M_{\rm dust}^{\rm CM} (M_\odot)$\tablenotemark{b}} & \colhead{$\log M_{\rm dust}^{\rm S} (M_\odot)$\tablenotemark{c}} & 
\colhead{$\log M_{\rm gas}^{\rm V} (M_\odot)$\tablenotemark{d}} & \colhead{$\log M_{\rm gas}^{\rm MS} (M_\odot)$\tablenotemark{e}}}
\startdata

5MUSES-179              & 0.45 & 0.11 & $7.23\pm0.10$ & $7.11\pm0.04$ & 8.81 & \, 9.82 \\
5MUSES-169              & 0.42 & 0.10 & $7.55\pm0.25$ & \nodata & 8.55 & \, 9.58 \\
5MUSES-105              & 0.26 & 0.06 & $7.49\pm0.63$ & $7.39\pm0.09$ & 8.60 & \, 9.85 \\
5MUSES-171              & 0.39 & 0.19 & $7.86\pm0.13$ & \nodata & 8.92 & \, 9.99 \\
5MUSES-229              & 0.21 & 0.09 & $7.86\pm0.66$ & $7.74\pm0.03$ & 8.87 & 10.21 \\
5MUSES-132              & 0.41 & 0.22 & $7.82\pm0.09$ & $7.83\pm0.03$ & 9.05 & 10.10 \\
5MUSES-227              & 0.24 & 0.02 & $6.65\pm0.77$ & $7.15\pm0.08$ & 8.38 & \, 9.66 \\
5MUSES-141              & 0.30 & 0.14 & $7.85\pm0.23$ & $7.65\pm0.04$ & 8.93 & 10.12 \\
5MUSES-136              & 0.29 & 0.15 & $7.67\pm0.14$ & $7.82\pm0.04$ & 9.14 & 10.33 \\
5MUSES-216              & 0.30 & 0.19 & $7.97\pm0.17$ & $7.89\pm0.02$ & 9.09 & 10.28 \\
PEPJ123646+621141       & 0.28 & 0.34 & $8.33\pm0.12$ & \nodata & 9.36 & 10.57 \\
GN70.38                 & 0.22 & 0.35 & $8.38\pm0.13$ & $8.27\pm0.49$ & 9.56 & 10.86\\
EGS13035123			    & 0.22 & 0.67 & $8.94\pm0.13$ & \nodata & 9.73 & 11.05 \\
PEPJ123759+621732       & 0.25 & 0.36 & $8.49\pm0.16$ & \nodata & 9.41 & 10.67 \\
GN70.8                  & 0.24 & 0.36 & $8.34\pm0.11$ & $8.64\pm0.24$ & 9.55 & 10.83\\
PEPJ123750+621600       & 0.24 & 0.33 & $8.24\pm0.11$ & \nodata & 9.55 & 10.82 \\
SMMJ163658+405728		& 0.25 & 0.74 & $8.85\pm0.09$ & $8.89\pm0.12$ & 9.75 & 11.01 \\
EGS13003805				& 0.21 & 0.43 & $8.33\pm0.08$ & \nodata & 9.86 & 11.20 \\
J123634.53+621241.3     & 0.20 & 0.67 & $8.72\pm0.08$ & $8.46\pm0.36$ & 9.98 & 11.35\\
GN70.104                & 0.25 & 0.42 & $8.38\pm0.09$ & $8.74\pm0.17$ & 9.63 & 10.90 \\
PEPJ123712+621753       & 0.35 & 0.19 & $7.63\pm0.08$ & \nodata & 9.21 & 10.32 \\
SMMJ030227+000653       & 0.24 & 0.64 & $8.63\pm0.07$ & $8.84\pm0.13$ & 9.82 & 11.09\\
EGS13004291			    & 0.18 & 0.57 & $8.63\pm0.09$ & \nodata & 9.99 & 11.39 \\
no.226                  & 0.24 & 0.26 & $8.12\pm0.13$ & $8.84\pm0.13$ & 9.44 & 10.71  \\
RG J123645.88+620754.2   & 0.24 & 0.42 & $8.24\pm0.07$ & $7.88\pm1.38$ & 9.78 & 11.07\\
3C298                   & 0.23 & 8.22 & $10.07\pm0.01$ & $9.51\pm0.04$ & 10.98\,\,\, & 12.29 \\
HR10                    & 0.10 & 0.31 & $8.96\pm0.63$ & $9.15\pm0.13$ & 9.80 & 11.48 \\
BzK-4171                & 0.21 & 0.34 & $8.20\pm0.09$ & $7.86\pm1.35$ & 9.73 & 11.06\\
51613					& 0.20 & 0.75 & $9.24\pm0.20$ & \nodata & 9.68 & 11.04 \\
BzK-21000			    & 0.10 & 0.08 & $7.95\pm0.84$ & $8.46\pm0.24$ & 9.38 & 11.05 \\
51858					& 0.26 & 0.55 & $8.80\pm0.14$ & \nodata & 9.51 & 10.76 \\
MIPS8342                & 0.10 & 0.44 & $8.16\pm0.05$ & $8.65\pm0.23$ & 10.67\,\,\, & 12.32 \\ 
3C318                   & 0.17 & 0.60 & $8.75\pm0.10$ & $9.47\pm0.11$ & 9.99 & 11.42 \\
SMMJ105151+572636       & 0.20 & 0.57 & $8.78\pm0.12$ & $8.85\pm0.16$ & 9.78 & 11.13 \\
COSB011                 & 0.18 & 0.75 & $8.69\pm0.06$ & \nodata & 10.18\,\,\, & 11.58 \\
SMMJ123555+620901       & 0.15 & 0.51 & $8.70\pm0.13$ & $8.92\pm0.15$ & 10.00\,\,\, & 11.48 \\
SMMJ123632+620800		& 0.08 & 0.15 & $8.54\pm1.05$ & $8.92\pm0.10$ & 9.70 & 11.47 \\
SMMJ123712+621322       & 0.22 & 0.74 & $8.67\pm0.06$ & $9.07\pm0.07$ & 10.00\,\,\, & 11.32\\
SMMJ123618.33+621550.5  & 0.23 & 0.94 & $8.84\pm0.05$ & $8.88\pm0.13$ & 10.06\,\,\, & 11.36 \\
SMMJ123711+621325       & 0.20 & 0.78 & $8.64\pm0.05$ & $9.20\pm0.06$ & 10.16\,\,\, & 11.52\\
RG J123711              & 0.30 & 1.07 & $8.82\pm0.04$ & \nodata & 9.91 & 11.09 \\
MIPS16080               & 0.22 & 0.56 & $8.53\pm0.07$ & $8.48\pm0.34$ & 9.86 & 11.17\\
SMMJ021738-050339       & 0.17 & 0.74 & $8.79\pm0.08$ & $8.84\pm0.17$ & 10.17\,\,\, & 11.60 \\
RGJ123644.13+621450.7   & 0.19 & 0.35 & $8.32\pm0.12$ & $8.94\pm0.09$ & 9.79 & 11.18 \\
HS1002+4400				& 0.38 & 5.99 & $9.55\pm0.01$ & $9.26\pm0.08$ & 10.64\,\,\, & 11.73 \\
MIPS15949               & 0.09 & 0.16 & $8.47\pm0.76$ & $8.73\pm0.18$ & 9.63 & 11.32\\
MIPS16144			  	& 0.19 & 1.11 & $9.21\pm0.09$ & \nodata & 10.07\,\,\, & 11.45 \\
MIPS429                 & 0.15 & 0.35 & $8.49\pm0.17$ & $8.63\pm0.24$ & 9.84 & 11.32\\
SMMJ123549+6215         & 0.08 & 0.28 & $9.03\pm0.91$ & $9.04\pm0.10$ & 9.79 & 11.54 \\
RXJ124913.86-055906.2   & 0.41 & 5.99 & $9.83\pm0.01$ & $9.04\pm0.08$ & 10.34\,\,\, & 11.38 \\
SMMJ021725-045934       & 0.20 & 0.72 & $8.82\pm0.08$ & $8.82\pm0.18$ & 9.97 & 11.34 \\
MIPS16059				& 0.18 & 0.33 & $8.40\pm0.16$ & \nodata & 9.70 & 11.11 \\ 
SMMJ16371+4053          & 0.37 & 2.24 & $9.16\pm0.02$ & $9.21\pm0.10$ & 10.10\,\,\, & 11.19 \\
SMMJ163650+405734       & 0.20 & 1.04 & $8.80\pm0.04$ & $9.12\pm0.08$ & 10.30\,\,\, & 11.67 \\
SMMJ105227+572512       & 0.25 & 0.72 & $8.76\pm0.08$ & $8.82\pm0.39$ & 9.78 & 11.04 \\
SMMJ16358+4105          & 0.10 & 0.28 & $8.53\pm0.29$ & $9.20\pm0.08$ & 10.02\,\,\, & 11.69 \\
SMMJ123707+6214         & 0.17 & 0.69 & $8.64\pm0.06$ & $8.85\pm0.14$ & 10.22\,\,\, & 11.65 \\
SMMJ123606+621047       & 0.20 & 0.43 & $8.55\pm0.13$ & \nodata & 9.71 & 11.07 \\
SMMJ221804+002154       & 0.42 & 5.99 & $9.83\pm0.01$ & $9.12\pm0.11$ & 10.33\,\,\, & 11.37 \\
SMMJ021738-050528       & 0.18 & 0.83 & $8.77\pm0.06$ & $9.02\pm0.09$ & 10.24\,\,\, & 11.65 \\
VCV J1409+5628			& 0.48 & 8.35 & $9.67\pm0.01$ & $9.61\pm0.02$ & 10.61\,\,\, & 11.59 \\
AMS12                   & 0.73 & 7.68 & $8.93\pm0.01$ & $9.15\pm0.07$ & 10.73\,\,\, & 11.53 \\
SMMJ04135+10277         & 0.50 & 11.57\,\,\, & $9.95\pm0.01$ & $9.56\pm0.05$ & 10.64\,\,\, & 11.60 \\
B3 J2330+3927           & 0.08 & 0.33 & $8.59\pm0.23$ & $9.33\pm0.05$ & 10.29\,\,\, & 12.04 \\
6C1909+722			    & 0.25 & 1.76 & $8.92\pm0.02$ & \nodata & 10.46\,\,\, & 11.73 \\
4C41.17                 & 0.08 & 0.25 & $8.50\pm0.34$ &$9.15\pm0.04$ & 10.15\,\,\, & 11.92 \\
GN20                    & 0.37 & 6.39 & $9.54\pm0.01$ & $9.22\pm0.07$ & 10.72\,\,\, & 11.81
\enddata
\tablenotetext{a}{Variable $\alpha_{\rm CO}$ calculated from \citet{narayanan2012}.}
\tablenotetext{b}{ISM radius, $R$, and $M_{\rm dust}^{\rm CM}$ are the parameters derived by fitting a radiative transfer model to far-IR photometry using the method of \citet{chakrabarti2008}.}
\tablenotetext{c}{$M_{\rm dust}^{\rm S}$ is calculated using the formalism of \citet{scoville2016}.}
\tablenotetext{d}{$M_{\rm gas}^{\rm V}$ is calculated using the variable $\alpha_{\rm CO}$ in Column 2. Errors are the same as for $L^\prime_{\rm CO}$ in Table \ref{properties}.}
\tablenotetext{e}{$M_{\rm gas}^{\rm MS}$ is calculated using a single $\alpha_{\rm CO}=4.6$ for every galaxy. Errors are the same as for $L^\prime_{\rm CO}$ in Table \ref{properties}.}
\end{deluxetable}
\bibliographystyle{aasjournal}

\end{document}